# News Recommender System: A review of recent progress, challenges, and opportunities

**Shaina Raza[1*], Chen Ding[1]**

**Abstract**: Nowadays, more and more news readers read news online where they have access to millions of news articles from multiple sources. In order to help users find the right and relevant content, news recommender systems (NRS) are developed to relieve the information overload problem and suggest news items that might be of interest for the news readers. In this paper, we highlight the major challenges faced by the NRS and identify the possible solutions from the state-of-the-art. Our discussion is divided into two parts. In the first part, we present an overview of the recommendation solutions, datasets, evaluation criteria beyond accuracy and recommendation platforms being used in the NRS. We also talk about two popular classes of models that have been successfully used in recent years. In the second part, we focus on the deep neural networks as solutions to build the NRS. Different from previous surveys, we study the effects of news recommendations on user behaviors and try to suggest possible remedies to mitigate those effects. By providing the state-of-the-art knowledge, this survey can help researchers and professional practitioners have a better understanding of the recent developments in news recommendation algorithms. In addition, this survey sheds light on the potential new directions.

✉ Shaina Raza*
Email: shaina.raza@ryerson.ca

Chen Ding
Email: cding@ryerson.ca

[1] Ryerson University, Toronto, Canada

# 1    Introduction

With the advancement in interactive communication technology, the internet has become a major source of news due to its 24/7 availability, instant updating and free distribution. According to a report by Pew Research Center Journalism in 2018[1], roughly nine-in-ten adults (93%) in US tend to read news online (either mobile or desktop) through digital newspapers, social media, news apps, etc. Despite such an advancement in technology, the studies have shown that online media does not define significantly different criteria for newsworthiness (Shoemaker 2006) than printed media. One reason for this could be the lack of prescribed procedures to offer a wide variety of news in a timely manner and the inability of the system to model user behaviors in a better way. Therefore, there is a need to move towards tools and techniques such as recommender systems (Adomavicius and Tuzhilin 2005) to provide news updates tailored to readers' information needs.

Many news sources and agencies such as CNN, BBC, New York Times, The Washington Post provide anytime, anywhere access to news readers so that they browse through latest news using online portals. To attract higher volume of traffic to their websites, these online portals are increasingly adopting recommender systems to improve user experience on their sites. The term 'user experience' may have different interpretations in a recommendation domain, such as usability, usefulness, effectiveness or satisfactory interaction with the system (Konstan and Riedl 2012; Knijnenburg et al. 2012). The task of recommending appropriate and relevant news stories to news readers is challenging. The reason is that the news domain is faced with certain challenges that are different from those of other application domains of recommender systems.

Among these unique challenges, timeliness is one of the most important challenges. It takes into account factors such as very short duration of news stories, their recency, popularity, trends, and a high magnitude of news stories arriving every second. Another important challenge in news domain is the highly dynamic user behavior. News readers may have long-term or short-term preferences that evolve over time, either gradually or abruptly. Recently, there is a considerable amount of manipulation taking place with the news content. For example, deceptive information is disseminated to the public in the form of false news and propaganda (Helberger 2019). This has given rise to an uprising challenge in terms of quality control of the news content.

As mobile technologies and applications become more prevalent in people's lives, news feeds from news aggregators (such as Google, Yahoo) and social media (such as Facebook and tweets) have taken over how people discover the news content. Once a news portal's recommendation functionality is installed, news feeds can be algorithmically tailored for each user. Personalization is a useful feature of NRS since it gives news based on the preferences and interests of a news reader. However, overly personalized news stories limit readers' exposure to different types of news. At the individual level, a news reader may get bored of reading similar types of news stories all the time. Over-personalization may also affect a reader's behavior in the long run, causing them to avoid counter-attitudinal (attitude that contradicts one's own beliefs) information (viewpoints, opinions)

---

(Helberger 2019). This type of behavior, at the societal level, poses a threat to democracy in the form of people's denial of opposing viewpoints.

Too much personalization in an NRS is often the result of recommendation approaches that place too much emphasis on prediction accuracy. These typical accuracy-centric approaches may fail to consider other aspects of subjective user experiences (such as choice satisfaction, perceived system effectiveness, better recommendations, and exposure to different points of view) when evaluating the recommendation quality. When developing a good NRS, one must consider the beyond-accuracy aspects to evaluate the quality of news recommendations.

### *Previous Surveys and Challenges Discussed*

In addition to the NRS-related papers, we also reviewed the previous surveys to see what they had covered. The challenges addressed in the literature often correspond to what is being investigated in the research during that time. For example, in classical NRS surveys (Borges and Lorena 2010; Karwa 2015; Dwivedi and Arya 2016), issues such as personalization, accuracy, cold-start problem, and scalability have been discussed. In some later NRS surveys (Karimi et al. 2018; Chakraborty et al. 2019), the new issues addressed (in addition to those covered in previous surveys) are beyond-accuracy aspects. Recently, the NRS surveys (Li and Wang 2019; Feng et al. 2020; Qin and Lu 2020) have covered topics such as cold start, news content and feature engineering, and changing user preferences. The challenges discussed by each of these surveys are listed in Table 1.

**Table 1**. Challenges discussed in different NRS surveys

| Survey | Challenges discussed |
|---|---|
| (Borges and Lorena 2010) | Accuracy |
| (Karwa 2015) | Cold-start, data sparsity, recency, implicit user feedback, changing interests of users, scalability, unstructured content |
| (Dwivedi and Arya 2016) | Data sparsity, changing users' interests, news content, recommendation techniques |
| (Karimi et al. 2018) | Recommendation paradigms, user modeling, cold start, data sparsity, recency, beyond-accuracy measures, scalability |
| (Chakraborty et al. 2019) | Recency, relevance, diversity, accuracy, recommendation techniques. |
| (Li and Wang 2019) | Recency, popularity, massive and unstructured data |
| (Qin and Lu 2020) | News content feature engineering. |
| (Feng et al. 2020) | Cold start, explicit user feedbacks, changing users' interests. |

Each of the preceding surveys revealed a few issues related to the news recommendation problem. However, their discussions are mostly from the perspective of computer scientists, ignoring the effects of news recommendation on user behaviors. Also, in the past few years, deep learning has become a popular option for building recommender systems, but they were not included in these surveys. Below we list the major differences between our paper and the previous surveys on the NRS.

1. In the previous surveys, the common challenges related to the news domain were considered. In addition to these common challenges, such as timeliness and user modelling, we discuss new challenges such as content quality and the effects of news recommendations on user behaviours. We provide an overview of the state-of-the-art research that addresses these new challenges.

2. We focus on the most popular recommendation models that are successfully used to build the NRS, with a special emphasis on deep learning-based models due to a lack of coverage on this topic in previous surveys.

3. The impact of news recommendations on user behaviours is a growing concern in the news industry. Although this issue has been raised by online journalism (Möller et al. 2018; Helberger 2019), we believe that it is also related to the discipline of computer science and information systems. Thus, different from previous surveys, we discuss changes in user behaviours that come in effect after recommendations. We also discuss possible remedies from computer science, psychology and journalism that do exist but have not been fully applied in recommender systems to mitigate those post-algorithmic news recommendation effects. In the discussion section, we also offer our own ideas of possible remedy approaches.

### *Searching Strategy, Scope and Research Trends*

In this survey, we have defined a searching strategy, scope, research goals and objectives to classify the literature. We take a neutral stance while reviewing the papers to avoid any risk of bias in the included studies. We identify and select the following collections of bibliographies: ACM Digital Library, SpringerLink, IEEE Xplore and Elsevier, to find the pertinent literature. Besides those bibliographies, we use the following scholarly search engines: GoogleScholar, DBLP, CiteSeerX, MS Academic Search, Web of Science, ScienceDirect and ResearchGate, to find the related papers. We also browse the conference proceedings and journal transactions to look for the titles and abstracts to find more papers which might have been initially skipped in the earlier search. We specify mid-year 2012 as the starting date and early 2021 as the closing date for our literature review. Besides this specified time frame, we also include a few classical and a few latest publications because of their relevancy to the topic.

We use the Boolean search query (("News") AND ("Recommender System" OR "Recommendation System" OR "Recommendations") OR (("Deep Learning) AND ("News Recommendations" OR "News Recommenders") to search the bibliographies with the following inclusion criteria: (i) papers written in English and (ii) relevancy and usefulness to the topic. Processing all the papers strictly was not practical. As a result, we decided to include only journal and conference papers, excluding grey literature, workshop presentations, and papers that report abstracts or presentation slides. Out of around 156 papers from the data extraction process, we finalize around 126 papers, out of which 92 are the manuscripts that proposed or designed an NRS, 8 are the survey papers and 26 are those that help us study the nature of the news domain. Out of the last group of papers, some articles are from journalism, general recommendation and information filtering domains. Figure 1 shows the approximate number of NRS papers considered in this time frame in a per-year basis.

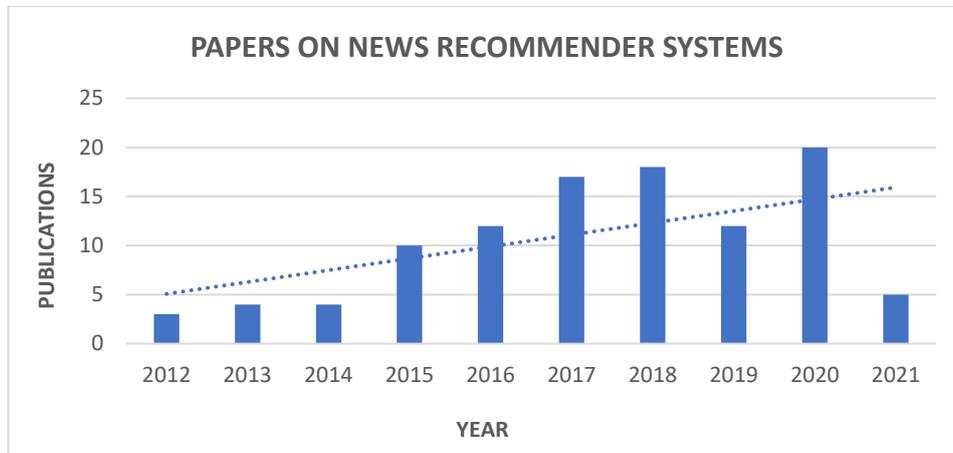

**Fig. 1.** Number of papers on NRS per year from mid-year 2012 till early year 2021

The figure clearly shows the increasing amount of research and demand for NRS in the field of recommender systems. The increase in the trendline in the later years is credited to the CLEF NEWSREEL Challenge (Brodt and Hopfgartner 2014) as well as the emergence and development of deep learning based recommender systems. The CLEF NEWSREEL platform (a campaign-style evaluation lab) was designed to encourage researchers to develop novel recommenders for news domains, so we see a clear rise in the number of publications during the years 2015-2017. Although it ended in 2018, due to the continuation pattern, we still see many papers in 2018. The effect of its ending is reflected in 2019, hence we see less work in 2019. Since 2016, there is a gradual increase of papers on deep learning-based recommender systems, both in the general domain and in the news domain. A higher number of publications in the year 2020 is possibly credited to the benchmark dataset MIND (from Microsoft). This trend is expected to continue in the year 2021, when the MIND dataset is released for the news recommendation challenge[2].

The primary goal of this survey paper is to highlight the most pressing challenges in the NRS that affect user behaviors at various stages of the news recommendation life cycle (before, during, and after).

The rest of the paper is organized as follows. In Section 2, we highlight the characteristics of the news domain. In Section 3, we present an overview of research on the NRS. In Section 4, we describe the conventional algorithmic solutions for addressing the major challenges in the NRS. In Section 5, we focus on the deep learning-based solutions to the NRS. In Section 6, we explain the effects of news algorithms on user behaviors. We discuss the research implications and future work in this field in Section 7. Finally, we conclude the survey in Section 8. A mind map diagram is given in Figure 2 that shows the evolution of this survey.

---

[2] https://msnews.github.io/competition.html

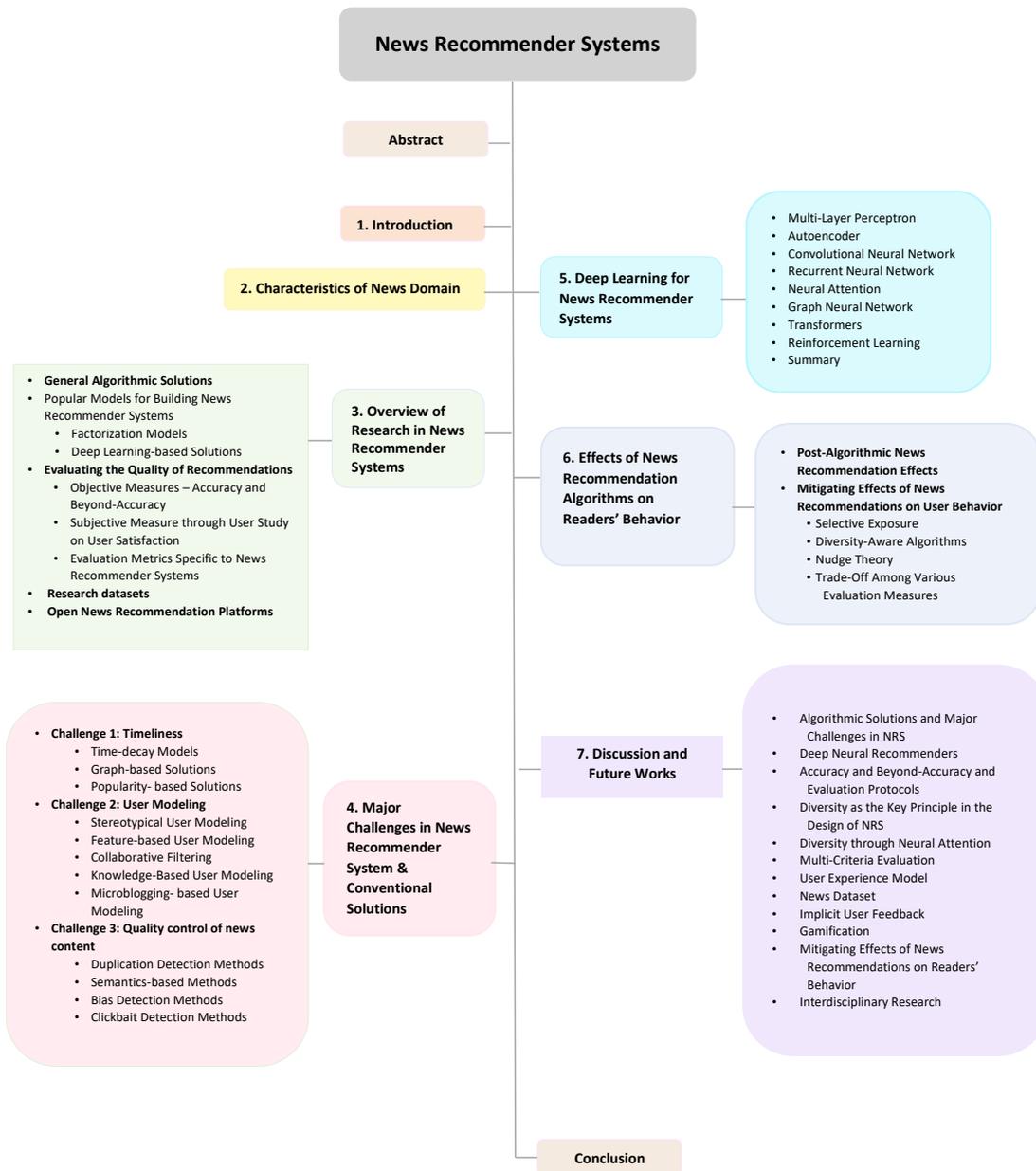

**Fig. 2.** Mind map diagram of the survey paper

## 2    Characteristics of News Domain

Before reviewing the challenges specific to NRS, we first highlight the characteristics that distinguish the news domain from other application domains of recommender systems such as recommending movies, music, books, restaurants or such.

**Average Consumption Time:** Typically, the duration of consuming a news story (time taken by a user to read a news article) is measured in terms of the article length which on average is under 200 words. According to a report by PEW research center[3], stories under 250 words require readers an average of 43 seconds in terms of the engagement time, whereas stories whose word count exceeds 5,000 engage people for at least 270 seconds (4.5 minutes). Compared to this, a movie is typically 90~120 minutes long, a music item on average is between 3~5 minutes long and a book may take even longer duration.

**Lifespan of News Items:** News items typically have shorter shelf-lives as they expire quite soon (maybe minutes, hours or barely a few days) compared to other products such as music, books, movies that may span several days, weeks, months or even years. Also, the gap between a news item's release time and time of reviews (comments) on news sites or social media sites is minimal (second, minute, hour or so) compared to other products.

**Catalog Size of News Items:** News stories tend to flood the system within a very short span of time, for example at the rate of thousands of incoming news items per hour. On the other hand, the catalog size of music or movie services may range typically in hundreds or thousands, but these items stay for longer time periods.

**Expected Request-Response Rate:** Timely delivery of the news content is vital and considered as a unique characteristic in the news domain. The requests for news items on a news aggregator site is sometimes greater than 100/sec and the expected response should preferably be sent within 100ms in order to provide news in real-time (Kille et al. 2017).

**Sequential Consumption:** News items are often consumed in a sequence where a reader may want to be updated about different news stories at a time. The difference between the sequential consumption of music items and news items is that in the former case the items are often repeated more than once within a sequence (Schedl et al. 2018), whereas in the latter case, the readers want to be updated with different or ongoing stories rather than the repeated stories (Park et al. 2017a).

**Diversity:** A user usually consumes one music or movie genre at a time and occasionally switches to a different genre when in a different mood or situation. On the other hand, the diversity in news domain is crucial not only to keep readers engaged during the online reading process but also to expose readers to counter-attitudinal behavior (Raza and Ding 2020). Diversity in news media is a key principle for a democratic society (Helberger 2019).

**Consumption Behavior:** News items are often consumed anonymously and mostly without explicit user profiles (Doychev et al. 2015; Sottocornola et al. 2018). Though this problem can be mitigated by considering implicit signals like click patterns, reading time spent on an item, browsing and navigational patterns (Ilievski and Roy 2013; Trevisiol et al. 2014), these implicit signals may sometimes be wrongly interpreted as an indicator of user's appreciation or interests. For example, longer reading time could be because of the user's fatigue or idle time and may not be an indicator of the user interest (Ma et al. 2016).

---

3    https://www.journalism.org/2016/05/05/2-analysis-suggests-some-readers-willingness-to-dig-into-long-form-news-on-cellphones/

**Privacy Concern:** Online media consumption has also resulted in the threat to users' privacy through excessive analyses on readers' data (Desarkar and Shinde 2014).

**Reading Context:** Reading context is highly evolving, time-ordered and social, and is specific to the news domain (Raza and Ding 2020). The most widely used contexts in NRS are location (Asikin and Wörndl 2014) and time (Park et al. 2017b). Lommatzsch et al. (Lommatzsch et al. 2017) evaluated users' dynamics with respect to the context of time and day of the week. According to their findings, there are more visitors on news portals during working days than on weekend. In addition to time and location, a reader's context may relate to some latest event or trending news, weather or even some personality trait (mood, interest). For example, during the Olympic games, people who are usually not interested in sports news may want to get update on the latest results of some games.

**Impact of Social Media:** Social media has greatly influenced the way news stories are searched and gathered (Cucchiarelli et al. 2018). Readers like to learn more about a news story by tracking its impact on social media. The dialogue, duration, public reactions and outcomes of a news story on social media may also help the journalists to determine which issues need further attention.

**Emotions:** Emotion commands attention and creates feelings in a reader for the event/character. A music or movie item intuitively evokes emotions in users, which in turn affect their preferences. Emotions are increasingly driving the news consumption behavior and they are both a challenge to the quality of what is produced, and also a chance for the NRS to further reinvent itself (Beckett and Deuze 2016).

**Biases:** News items are initially consumed for information purpose; however, biases can be invoked through presenting news in different styles and tones (Helberger 2019). A good news story should be one that offers details to the readers so that they can make their own judgement and forge an emotional connection with a character/event.

**Multimodal News Information:** In today's information age, the Web is critical for disseminating information and news. Social media, in particular, can easily notify users of global events and has grown in popularity as a big source of news. These news articles often use multiple modalities, such as texts, videos, podcasts, to convey information more effectively. When it is in the text format, it can be delivered in different languages. Most of the research work today focus on the text-based news articles in one language, without considering the complications brought by multiple modalities and languages, since it is challenging to quantify the cross-modal and cross-language entity representations in today's news domain. Due to a lack of active research on recommending news in the non-textual format and in multiple languages, in this survey, we only review the papers on recommending text-based news in one language. However, we do recognize the need to have more research on multi-modal and multi-language news recommendation.

## 3  Overview of Research in News Recommender Systems

We present an overview of the NRS research in this section. In Section 4, we present the major challenges for the NRS and some conventional solutions to address them. In Section 5, we present the deep learning-based NRS.

### 3.1 General Algorithmic Solutions

The traditional algorithms used in recommender systems can be classified as: collaborative filtering (CF), content-based filtering (CBF) and hybrid approaches (Adomavicius and Tuzhilin 2005). There are two important things required to build any recommender systems, i.e., the content of the users and items, and their underlying interactions. A CBF algorithm builds a recommender by comparing the user-profile and item-profile based on the content of a shared attribute space. Contrary to this, the CF approach is content-free where the features of items are often not known in prior. CF exploits user behaviors in terms of ratings, history and interactions on items.

While these traditional recommendation algorithms can be applied to the news domain, their performance may not be good. There are various scenarios that we need to consider, such as the dynamics of the news environment, relevance of the news items and users' interests that are highly context dependent. Though CF can be used to address the problem of dynamic content generation of news items, it requires a sufficient amount of users' interactions (stored as histories) to make recommendations. By the time NRS manages to collect enough consumption data from users, the value of news content is decayed, thus making recommendations obsolete. The CBF, on the contrary, can address users' evolving interests by always updating the user profiles with the latest news they have read (Wang et al. 2018b). However, CBF cannot handle the large number of temporary and anonymous users that are common in an NRS. Also, the statistical methods to compute the similarity between user-item profiles in a CBF, may fail to capture the semantics and the contexts in news data. To remedy the pitfalls of both CF and CBF in the NRS, researchers and designers propose hybrid solutions to news recommendations by combining these two types of algorithms. In the past few years, researchers also began to focus on the context (situation such as time, location, mood, etc.) as the additional information to improve the quality of news recommendations. An analysis of 79 (out of 92) papers on NRS in our survey is shown in Figure 3.

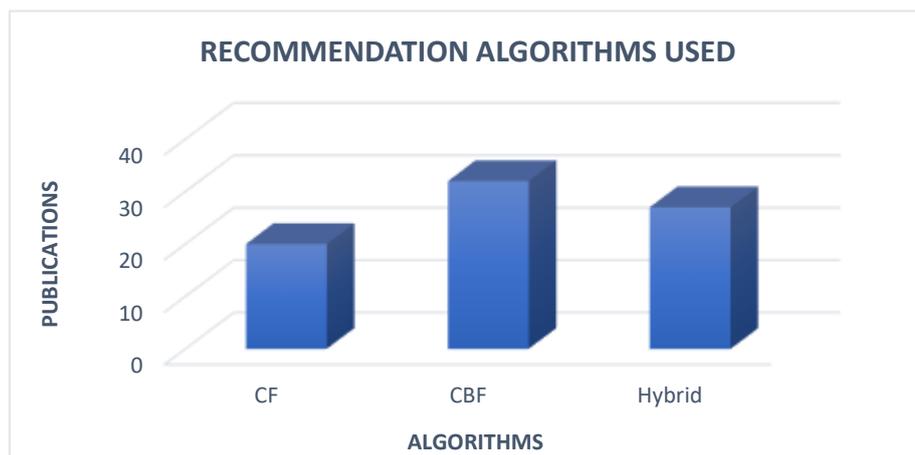

**Fig. 3**. Algorithms used in NRS

The statistics in Figure 3 show that CBF is the most used algorithm to solve the problem of news recommendations. Since CBF methods are primarily based on the content metadata to produce recommendations, it is much easier for the researchers and developers to develop an NRS. The hybrid comes up as the second most popular choice to build an NRS. The CF comes as the next popular choice (also the least popular among the three).

### 3.2 Popular Models for Building News Recommender Systems

Many models have been used in the past to build an NRS. One of the most popular and successful classes of models for the NRS is the latent factor model, especially factorization methods. In recent years, the deep learning-based solutions have come up as an emerging branch of recommender systems. We consider them as the other most popular class of models successfully used for the NRS. These models are briefly covered below.

#### 3.2.1 Factorization Models

Factorization method is a class of algorithms used in recommender systems that work by decomposing the user-item interaction matrix into a product of lower dimensional matrices. Here we discuss the factorization models used in the NRS research.

***Matrix Factorization (MF)***
Matrix factorization is one of the most popular recommendation algorithms that got its first recognition in the Netflix competition (Koren et al. 2009). Matrix factorization can be used to discover the latent features that exhibit in the interactions between two different types of entities (e.g., users and items). In a recent NRS (Raza and Ding 2019), the MF is extended to include the news-related information and to model the temporal dynamics in readers' behaviors. This work introduces a novel predictor to include various temporal effects in the MF model, including time bias, user bias, and item bias. These added biases tend to capture much of the observed signals, especially the temporal dynamics.

***Non-Negative Matrix Factorization (NMF)***
NMF, like the MF method, is a decomposition technique in which the matrix R is split into the product of two matrices U and V. However, unlike MF, the NMF has the property that none of the three matrices R, U, and V have any negative elements. Typically, there are many missing user-item interactions in an NRS, resulting in very sparse matrices. In such situations, the NMF models usually perform better than the original MF. This is due to the default functionality of the NMF algorithm in addressing the missing-value assumption (Gillis 2020). However, Singular Value Decomposition (SVD)-based MF may produce better results if the ratings matrix is not too sparse.

In a related NRS (Yan et al. 2012), news-related information is included into the NMF model, where NMF is used for clustering news documents and topic discovery. In another paper (Shu et al. 2019), the NMF is used to learn latent space embeddings from news content and user-news interactions.

***Tensor Factorization (TF)***
TF extends the MF model by introducing the latent vectors with additional dimension(s). TF-based recommender systems go beyond the limitations of MF techniques by considering additional information about users and items, which results in more accurate recommendations (Frolov and Oseledets 2017). So, TF methods are useful in NRS scenarios where we need to consider more contextual recommendations, such as time, location, and social interactions. However, including too many dimensions may result in costly computations.

In a related NRS (Wang et al. 2015), TF is used to include the contextualized information related to news items and news readers into the recommendation model.

### Probabilistic Matrix Factorization (PMF)

PMF (Mnih and Salakhutdinov 2007) is a type of MF models with Gaussian observation noise. PMF is a variation of MF methods that takes its intuition from the Bayesian learning for parameter estimation. This model scales linearly with the number of observations and performs well on large, sparse, and highly imbalanced datasets like those found in the news domain.

In a social recommender system, PMF is used to combine social network structure and user-item rating matrix (Ma et al. 2008). The same idea is used in an NRS (Lin et al. 2012), which incorporates news content, user interactions, and social network information into a PMF model to address the data sparsity issue.

### Bayesian Personalized Ranking (BPR)

A general limitation of the traditional item prediction methods (e.g., MF methods) is that they are not optimized for ranking the items (e.g., news items). The BPR optimization uses pairs of items to produce more personalized rankings for each user. The MF models can also be used with the BPR to provide users with a personalized and ranked list of items (Rendle et al. 2012).

In a related NRS (Xia et al. 2014) based on a Bayesian model, the readers are recommended with the latest news stories by calculating the joint probability of the news. In another NRS (Gharahighehi and Vens 2019), an extension of BPR is proposed that uses the user consumption levels to recommend news topics to the readers.

### Generalized Linear Modeling (GLM)

A CF method is often formulated for the prediction of unobserved ratings in a large and mostly empty rating matrix. Though not strictly an MF methodology, both methods (MF and GLM) have their origins in latent factor models. The GLM (McCullagh 2019) can also be used together in conjunction with the CF, where it can use its probabilistic modelling to factorize a high-dimensional rating matrix. In a recent NRS (Raza and Ding 2020), the knowledge is transferred from a high-dimensional news domain that is factorized using GLM into a CF model. The CF model is then used to predict and recommend news items for the users.

### Neural Extensions

Much of the recent research in recommender systems is based on creating the neural extensions of these successful latent factor methods discussed above. For example, Neural Network Matrix Factorization (NNMF) (Dziugaite and Roy 2015) replaces the inner product in the PMF formulation with a neural network, and is able to learn an appropriate nonlinear function of user and item latent variables. Neural Collaborative Filtering (NCF) (He et al. 2017) extends the CF model, and the Deep Matrix Factorization (DFM) (Xue et al. 2017) extends the traditional MF model to map the users and items into a common low-dimensional space with non-linear projections. These models continue to inspire NRS researchers, resulting in several useful news recommendation models.

### 3.2.2 Deep Learning-Based Solutions

The deep learning based NRS began to evolve in the later years, i.e., since 2016 (Karatzoglou et al. 2016). In our survey, we found more than 30 papers published since 2017 that use deep neural networks to solve the news recommendation problem. The rising popularity of these methods shows that deep learning will become the most popular methods in the near future to work in this domain. The general statistics of deep learning based NRS are shown in Figure 4.

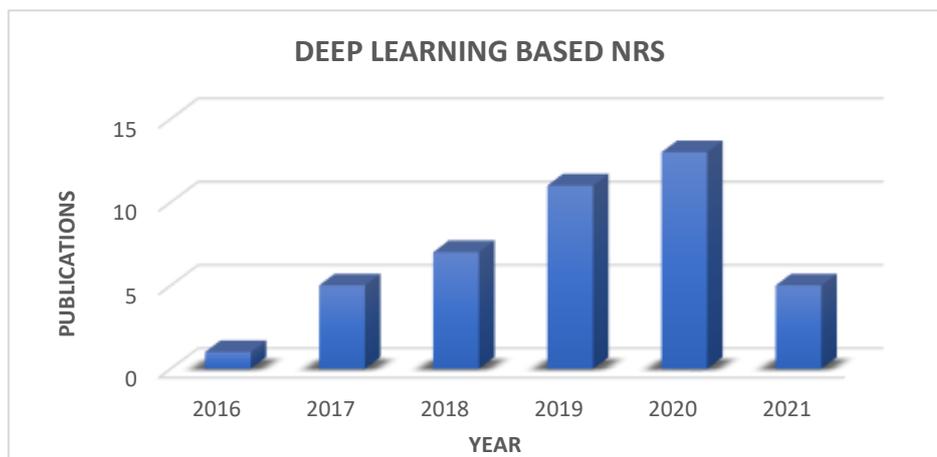

**Fig. 4**. Deep learning algorithms used in NRS from 2016 till early year 2021

As shown in Figure 4, the approach to deep learning is increasingly being employed to develop NRS solutions with every coming year. The number is lower in 2021, mainly because it is only the middle of 2021 at the time of writing of this paper and many papers are not yet published or posted online. We will go over the deep learning-based models to news recommendations in Section 5.

### 3.3 Evaluating the Quality of Recommendations

We categorize the evaluation measures in NRS into two types: objective measures – accuracy and beyond-accuracy, and subjective measures through the user study on user satisfaction. Below we review measures under each category and how they are used in different research work. The definitions for the actual evaluation metrics that have been used in NRS so far and which categories they belong to are given in Table 2.

### 3.3.1 Objective Measures – Accuracy and Beyond-Accuracy

The goal of a recommender system is to predict how likely users would enjoy the unknown items based on what the system has known about them. Therefore, much of the early work in recommender systems focused on providing recommendations to the users according to their preferences. These recommender systems have been evaluated according to accuracy metrics that measure the algorithm performance by comparing its prediction against a known user rating of an item (Herlocker et al. 2004; Gunawardana and Shani 2009). However, such accuracy-centric evaluations cannot answer the question about if users are satisfied with the recommendations. For example, Amazon claimed to generate an additional 10% to 30% of its revenue in 2015 from the sale of diverse (non-personalized) items (Srihari 2015). This kind of insufficiency has shifted some researchers' focus to different goals for a recommender system, which can address other aspects beyond accuracy. Generally, recommending everything related to users' preferences would result in good accuracy. However, for news consumption, though accuracy is important, other factors are equally crucial to satisfy users' needs. Below we discuss the beyond-accuracy aspects in NRS.

### Diversity

Diversity measures the degree of 'dissimilarity' among the recommended items. It is mostly implemented through re-ranking of the recommendation lists. Some well-known metrics are: Intra-List Similarity (ILS) (similarity between any two lists of recommended items); temporal or Lathia's diversity (in the sequence of recommendation lists over time); normalized diversity; and other measures as discussed by Kunaver and Porl (Kunaver and Porl 2017). The traditional pairwise diversity ILS remains a popular metric to evaluate diversity in NRS (Li and Li 2013; Gu et al. 2014; Maksai et al. 2015; Raza and Ding 2020). The ILS can be computed among the items, topics, categories, tags or even sentiments (tone) (Helberger 2019) in an NRS. Since the typical ILS method is computed for each individual user, it is a computationally expensive process for an NRS where there are millions of users and items. Thus, it requires more research to consider various aspects, such as level of diversification, scalability issues in an NRS.

### Coverage

Coverage represents the percentage of distinct items/users/ratings that a recommender system can recommend. Popular interpretations of coverage include item coverage (percentage of items), user coverage (percentage of users), catalog coverage (percentage of recommended user-item pairs) and interaction coverage (rating predictions) over potential items, users, user-item pairs, or ratings respectively (Han and Yamana 2017). Coverage in an NRS is treated no differently than that in other recommendation domains. It is mostly used to determine the coverage of items in the news domain (De Francisci Morales et al. 2012; Maksai et al. 2015). In some cases, coverage is defined as the measure of number of users' visits on the website during different times to determine the topical coverage [11]. The research on coverage for NRS is still very limited and discusses item coverage mostly. It is important to have more research on coverage since this aspect is related not only to the recommended items but also to the whole NRS.

### Novelty

Novelty determines how different or unknown a recommendation is to what has been previously recommended to a user (Vargas and Castells 2011). Silveira et al. (Silveira et al. 2019) defined novelty at three levels: a user has never heard of the item in his life (life-level), item is unknown to the user as per his consumption history (system level), and finally the non-redundant item in the recommendation list (recommendation level). Introducing novelty is more challenging in the NRS because almost everything that is happening inside the news domain is novel. In its simplest form, novelty is defined as the inverse of popularity or the ratio of unknown items in the top-N recommended list of news items (Garcin and Faltings 2013; Gu et al. 2014; Maksai et al. 2015; Saranya and Sudha Sadasivam 2017; Raza and Ding 2020). So far, the inclusion of novelty in NRS is limited to the item level only. Novelty should also be covered in terms of overall content, events and uniqueness of news stories to the users.

### *Serendipity*

Serendipity is a composite concept that includes various aspects such as the degree of relevance (usefulness), novelty (new) and unexpectedness (surprise) (Kotkov et al. 2016). Serendipity is different from novelty. An item is novel if a user is not familiar with or has not consumed or forgotten about the item, whereas an item is serendipitous if the user does not expect or would not have discovered this item but found it fortunate and interesting to have it recommended to him. For example, if the user is recommended a news story that he has never heard of, this news story is novel to him but not serendipitous if he is not interested in that topic. On the contrary, if the user finds this news story interesting enough to change his attitude on that news category or topic, this news item is a serendipitous item (Asikin and Wörndl 2014). In one NRS (Maksai et al. 2015), the serendipity aspect is defined as being composed of accuracy, novelty and diversity. In a few other NRS (Jenders et al. 2015; Cucchiarelli et al. 2018), serendipity is defined in terms of news topics that are both semantically related and are unexpected. The literature shows only limited research on the serendipity in the NRS. One reason for this could be that serendipity is a composite aspect with many combinatory definitions, which makes it difficult for the researchers to evaluate.

### 3.3.2    Subjective Measures through User Study on User Satisfaction

User experience is a subjective term, with different meanings and interpretations. It is affected by many factors during different stages of recommendations, i.e., before, during and after the recommendations are made. For example, recommending something trending or related to the user's context (e.g., demographics) during the sign-up process increases the user's loyalty to the system. Similarly, proactively recommending some news stories at the side pane during the normal reading process may persuade users to stay in the system longer. If a recommender system can include these features, it may increase the user's trust with the system.

In recommender systems, the user experience is usually evaluated through three prominent ways: (i) by carrying out user studies where the subjects are given certain questionnaires during different stages of recommendations (Konstan and Riedl 2012), (ii) by combining study on longitudinally logged data with the questionnaire-based user study (Nguyen et al. 2014), and (iii) by addressing other evaluation measures such as combining accuracy and beyond-accuracy measures in certain ways (Maksai et al. 2015).

The user experience framework by Knijnenburg et al. (Knijnenburg et al. 2012) consists of six components: objective system aspects (algorithm, presentation, interface and additional features of a recommender system); user experience (choices, evaluations of the system by the user); perception or subjective system aspects (user's evaluation of the objective aspects); situational (different contexts such as social, trust, choice goal) and personal characteristics (gender, location) as external features; and objective interaction (observable behavior such as browsing, viewing, signing-in, rating, consuming).

A few small-scaled frameworks for user experience are also proposed specifically for the NRS. The framework in one NRS (Asikin and Wörndl 2014) considers only three factors (i.e., appropriate, like, surprising) to evaluate user behaviors. The other framework (Constantinides and Dowell 2018) considers six factors, i.e., reading frequency, reading time, time of day, reading style, browsing strategy and location (context) to evaluate the user experience. However, these are all implicit factors and can only be used as indicators of user experience. They are not the direct measures for user experience.

The researchers in prior work in NRS have associated user satisfaction with objective measures. They assumed that user experience is a global phenomenon for all users, so they use one measure for all. For some researchers, it can be measured through accuracy (Nguyen et al. 2014; Viana and Soares 2016; Su et al. 2016). They demonstrated that higher ratings provide more pleasing and satisfactory experience to the users. For some other researchers, the user experience is more related to beyond-accuracy aspects. For example, a few authors (Asikin and Wörndl 2014; Jenders et al. 2015) claim that increasing serendipity in the NRS would yield higher user satisfaction. Some associate user experience with a higher degree of novelty (Saranya and Sudha Sadasivam 2017).

**Table 2.** Evaluation Metrics accuracy (acc): beyond-accuracy (beyond-acc)

| Metric | Description | Type |
|---|---|---|
| Accuracy | Number of correct predictions over total predictions | acc |
| Precision (prec) | Proportion of relevant items over total recommendations. | acc |
| Recall (rec) | Proportion of relevant items over total relevant items. | acc |
| F1-score (F1) | Weighted average of the precision and recall. | acc |
| Mean Reciprocal Rank (MRR) | Multiplicative inverse of rank of the first correct item. | ranking, acc |
| Mean Average Precision (MAP) | The average precisions across all relevant queries. | ranking, acc |
| **Rank** | Percentile-ranking within the ordered list items. | ranking, acc |
| Cumulative Rating | Total relevance of all documents above a rank position in top recommended items. | ranking, acc |
| Success | Current item that is in a set of recommended items. | ranking, acc |
| Novelty (nov) | Ratio of unseen items over recommended items. | beyond-acc |
| Serendipity (seren) | Measure of unexpectedness with relevance (relevance is 1 if recommended item is interacted with otherwise 0 ) | beyond-acc |
| Coverage (cov) | Percent of items that the model is able to recommend. | beyond-acc |
| Diversity (div) | Degree of dissimilar recommendations either at system level (aggregate diversity) or for each user (individual diversity). | beyond-acc |
| Hit Rate (HR) | Ratio of hits in ranked items over the number of users. | acc |
| Log-loss | Probability of a prediction input between 0 and 1. | acc |
| Average Reciprocal Hit Rate (ARHR) | A hit is inversely relative to its position in top recommendations. | acc |
| Root-mean-square error (RMSE) | Difference between predicted and actual rating. | acc |
| Click-through rate (CTR) | The likelihood of a news item that will be clicked. | acc |
| Discounted Cumulative Gain (DCG) | Gain of an item as per to its position in recommendation list. | acc |
| Area under curve (AUC) | ROC curve plots recall against fallout (false positive rate). | acc |
| Customer Satisfaction Index (CSI) | The satisfaction degree of user on the recommendations. | beyond-acc |
| Personalized (NRS-specific) | Current item that is in a set of recommended list but is not among the popular items (Garcin et al. 2013). | relevancy, acc |
| Saliency (NRS-specific) | A function of entities' frequency in news articles, with a decay factor (Cucchiarelli et al. 2018). | beyond-acc |
| Future-Impact (NRS-specific) | Evaluate a tradeoff between recency and relevancy (Chakraborty et al. 2019). | beyond-acc |
| Tradeoff (NRS-specific) | Evaluate a tradeoff between high accuracy and reasonable diversity (Raza and Ding 2020). | beyond-acc |
| Senti (NRS-specific) | Evaluate sentiment diversity of recommendations, motivated by MRR and hit ratio (Wu et al. 2020a). | beyond-acc |

We show a distribution of accuracy and beyond-accuracy metrics used in NRS papers in Figure 5.

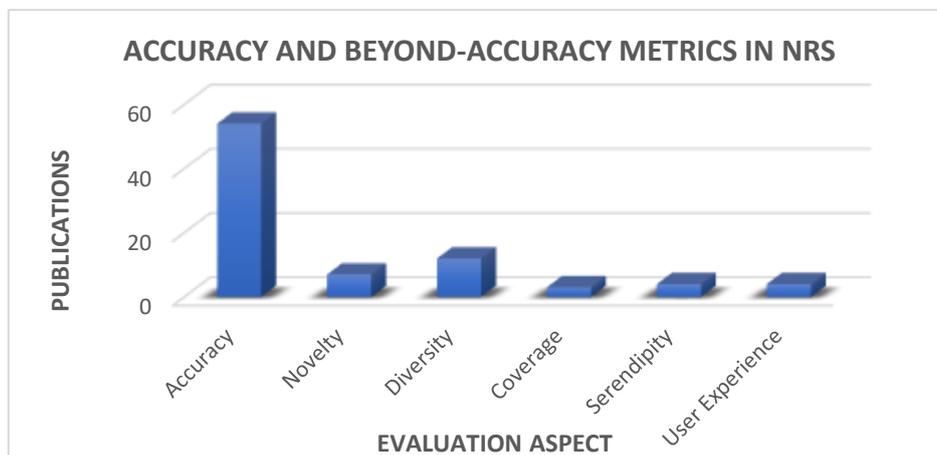

**Fig. 5.** Distribution of accuracy and beyond-accuracy aspects in NRS

The statistics in Figure 5 show that accuracy is the most widely used evaluation measure in the NRS. The researchers also put some efforts to introduce diversity in news recommendations. There is very limited work in novelty, coverage and the most important aspect, i.e., user experience in NRS. In general, the quality metrics used in the NRS research are more or less same as those used in general recommender systems. However, a few evaluation metrics are designed specifically for the NRS, which are discussed next.

### 3.3.3 Evaluation Metrics Specific to News Recommender Systems

#### *Personalized*
Garcin et al. (Garcin et al. 2013) propose a personalized @k metric, which removes *k* popular items from the recommendation list to produce a smaller set of recommendations. The goal is to eliminate the popularity bias that occurs when data is collected from websites that automatically recommend the most popular items.

#### *Saliency*
Cucchiarelli et al. (Cucchiarelli et al. 2018) propose a saliency metric. The saliency of entities (named entities) is calculated as a function of their frequency in news articles, with a decay factor based on the distance of the positional index of the first occurrence in the text. The idea of this metric is inspired by the news-specific discourse structure, which tends to provide brief summaries of the most important facts and entities in the first paragraphs.

Future-Impact: Chakraborty et al. (Chakraborty et al. 2019) propose a future-impact metric that tradeoffs between recency (age of a news story after it is published) and importance (relevance). A news story with a higher future-impact score is thought to be a high-impact story, and vice versa. Usually, the news stories that are recently published are given the highest future-impact scores.

*Tradeoff*

Raza and Ding (Raza and Ding 2020) propose a tradeoff metric for balancing high accuracy (precision, recall measures) with reasonable diversity (diversity and novelty aspects). The assumption is that higher accuracy leads to better personalization and thus improves readers' experiences with the NRS. Reasonable diversity, on the other hand, helps readers get diversified news so that they don't get bored reading the same news stories over and over. This metric is designed to keep readers engaged in the reading process while also recommending diverse news to them.

*Senti*

Wu et al. (Wu et al. 2020a) propose a metric Senti (from word sentiment) to evaluate the sentiment diversity of news recommendations. This metric normalizes MRR and hit ratio scores. The Senti is positive if the top-ranked news has the same sentiment orientation as the overall sentiment, and it is higher if the sentiments are stronger.

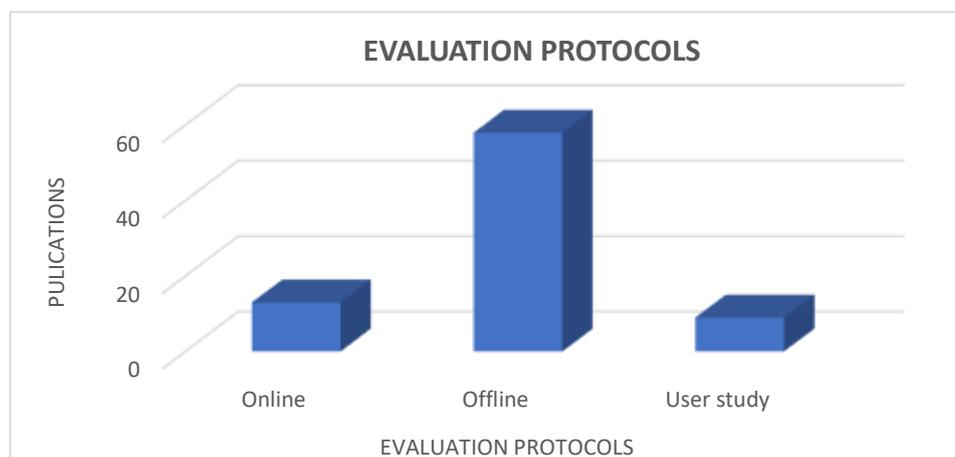

**Fig. 6.** Distribution of evaluation approaches used in NRS

We have also considered three standard evaluation protocols while classifying the literature for evaluation measures in Figure 6. These evaluation protocols refer to the experimental settings in which we measure the quality of recommendations and include offline experimentation/simulation, online experimentation (A/B or real-time tests) and user studies (Gunawardana and Shani 2009).

As can be seen in Figure 6, there are 13 papers using online evaluation, 58 using offline evaluation, and 9 using the user study. The offline evaluation protocol is the most widely used in the NRS research. One reason for this could be that online evaluation and the user study are often considered as an expensive approach in real-time settings in an NRS.

### 3.4 Research Datasets

Since the objects to recommend in the news domain are mostly text documents, the news datasets mainly consist of textual data. There are different types of datasets that we can consider: (i) publicly available datasets for non-commercial and research purposes, (ii) proprietary datasets, (iii) crawled datasets, or (iv) synthetic datasets created with simulated (anonymized or hidden or added) values. The details of a few datasets such as Plista, Adressa, Yahoo, Outbrain, and a few open source frameworks have been given by Karimi et al. (Karimi et al. 2018), so here we just give a brief overview of them. However, we also discuss some datasets that are either new or less discussed, such as Yahoo news, Hacker news, BuzzFeed and some fake news datasets.

### Plista

Plista is a dataset developed by Plista (an advertising company) and Technische Universität Berlin to promote research in NRS (Kille et al. 2013). It consists of logs from 13 German news portals collected from June 2013. It also contains millions of impressions (articles views) and some time-related information. This dataset is accessible upon request for research purposes.

### Adressa

Adressa (Gulla et al. 2017) is a publicly available benchmark dataset developed by Adressavisen (a local newspaper in Norway) and Norwegian University of Science and Technology (NTNU). Like Plista, Adressa does not have explicit ratings, but different from Plista, it includes reading time in addition to reading counts.

### Yahoo Webscope

Yahoo Webscope[4] is a reference library that provides datasets for non-commercial users such as academics and scientists. Yahoo provides benchmark datasets for news as well. These datasets are: R6A - *Yahoo! Front Page Today Module User Click Log Dataset*, R6B - *Yahoo! Front Page Today Module User Click Log Dataset*, R11 - *Yahoo News Video dataset*, L33 - *Yahoo News Ranked Multi-label Corpus* and L32 - *The Yahoo News Annotated Comments Corpus*. Among these datasets, the two news datasets (R6A and R6B) with ratings and news category information provided by Yahoo! Front Page Today are of importance for researchers to evaluate their recommendation algorithms. These two datasets consist of the timestamp information and explicit ratings, which makes them a favorite option for developing and evaluating CF solutions. However, one limitation of these datasets is that news items are represented by their features where actual content of the news stories are anonymized without any additional information. It might be difficult to make recommendations in the absence of any information on the stories. These datasets are also available upon request for research purposes.

### Hacker News

Hacker News[5], run by YCombinator[6], is a popular social news website. It is widely known among the people in the IT industry where they can share news, demonstrate their projects, ask questions, post jobs and comment on news stories as a community. Hacker News provides a big dataset under the MIT License since its launch in 2006. This dataset is also available as a public dataset through Google BigQuery[7] (a RESTFUL web service providing exploratory analysis of massive datasets in conjunction with Google storage).

---

[4] https://webscope.sandbox.yahoo.com/
[5] https://github.com/HackerNews/API
[6] https://news.ycombinator.com/
[7] https://cloud.google.com/bigquery/what-is-bigquery

This dataset consists of news stories from various sources, which may be useful for the researchers working on news recommendations. However, the texts and the comments on news do not go through the censor process and may include profanity. Hacker News does not take responsibility for what the authors have written.

*BuzzFeed News*

BuzzFeed[8] is a company that provides news and entertainment content on digital media. They publish data related to fake news, social media and various news patterns. They have released some datasets and made them available on GitHub[9]. These datasets are useful for researchers working on fake news investigating rumors, misinformation and detecting factual claims. However, one limitation is that these datasets are particularly for fake news detection and may not be a proper source for building a personalized NRS.

*MIcrosoft News Dataset (MIND)*

The MIND dataset is a large-scale benchmark dataset (Wu et al. 2020b) for news recommendation research. MIND contains about 160k English news articles, and more than 15 million impression logs generated by 1 million users. Every news article is identified by rich textual content including title, abstract, body, category, and entities. The impression log contains the click events, non-clicked events and historical news click behaviors of a user. MIND-small is a small version of the original MIND dataset, which consists of 50,000 users and their behavior logs. The users are anonymized. Both versions of the dataset can be accessed online[10].

*Fake News Datasets*

Fake news has become a serious problem for spreading rumors and misinformation, and consequently made negative impact on politics, regional stableness, and sometimes even people's daily life, especially during the US election and the pandemic period. Because of this, many fake news datasets are made accessible for open research in recent years. Though they are not directly related to the NRS research, they are useful for fake news detection, which could be a crucial step before making recommendations. A few prominent ones are listed here: BS Detector[11], Credbank-data[12,] BuzzFace[13], MisInfoText[14], NewsTrust[15], SFU Opinion and Comments Corpus[16], NELA-GT-2018 (Nørregaard et al. 2019), , NELA-GT-2019 (Horne 2020), NELA-GT-2020 (Gruppi et al. 2021), Fakeddit (Nakamura et al. 2019), FakeNewsNet (Shu et al. 2018), NYtimes covid-19-data[17] and LIAR (Wang 2017) datasets.

---

[8] https://www.buzzfeed.com/
[9] https://github.com/BuzzFeedNews/everything
[10] https://msnews.github.io/
[11] https://github.com/thiagovas/bs-detector-dataset
[12] http://compsocial.github.io/CREDBANK-data/
[13] https://github.com/gsantia/BuzzFace
[14] https://github.com/sfu-discourse-lab/MisInfoText
[15] http://resources.mpi-inf.mpg.de/impact/credibilityanalysis/data.tar.gz
[16] https://github.com/sfu-discourse-lab/SOCC
[17] https://github.com/nytimes/covid-19-data

*Other Datasets*

There are some classical news datasets such as Reuters Corpora[18] and 20 Newsgroups[19] that are used for news categorization. Some of the recent ones include Amazon news datasets at Fast.ai[20] and Global Database of Events, Language and Tone (GDELT)[21] that can be used for text categorization and detailed analyses on news and user data. Some news-related datasets[22] have recently been made public by the Huggingface library (a library for Transformer models).

In Figure 7, we show the distribution of datasets used in previous NRS research.

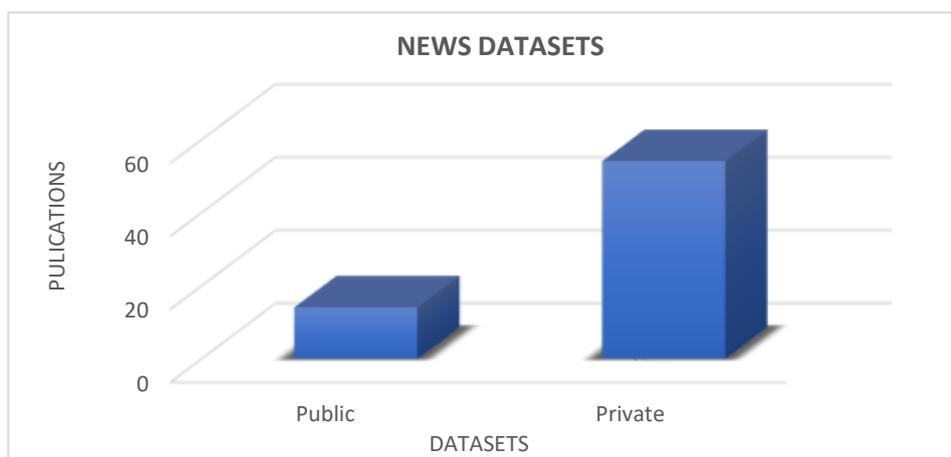

**Fig. 7.** Distribution of datasets in NRS

As can be seen in Figure 7, 62 papers are using private (mostly crawled) datasets, 16 papers are using public datasets.

Most of the time, the researchers prefer to build their own news recommendation datasets for two important reasons: lack of publicly available datasets and unique requirements on certain types of data for their research. In that, they crawl news from different news publishers. These datasets are usually proprietary to the organization who created them. There are also synthetic datasets that are domain dependent and are created by taking data from some benchmark datasets and enriching them by including related information and interactions either artificially or in a semi-autonomous way.

We have also included the information in Table 3 and Table 4.

### 3.5    Open News Recommendation Platforms

Over the last few years, there are several libraries that have been developed for recommendations. A few prominent ones are discussed briefly here.

---

[18] https://trec.nist.gov/data/reuters/reuters.html
[19] http://qwone.com/~jason/20Newsgroups/
[20] https://course.fast.ai/datasets
[21] https://www.gdeltproject.org/data.html
[22] https://huggingface.co/datasets?search=news

**MIND** (Wu et al. 2020b) is a recent news benchmark dataset. The contributors of the dataset provided an environment in the form of a competitive event and Leaderboard[23] for researchers to work on the news recommendation problem. In conjunction with The Web Conference 2021[24], the contributors also offered an International Workshop on News Recommendation and Intelligence. This incentive requested research and technical report articles on many aspects of news recommendations.

**Apache Mahout** is a distributed machine learning library implemented in Java and contains some CF algorithms. This framework is available both for academic and commercial use to work with real-world news data (Beck et al. 2017).

**Idomaar** (Scriminaci et al. 2016) is a benchmark framework that enables efficient reproducible evaluation of recommendation algorithms in real-world settings. Unlike other frameworks implemented in Java, Python or C++, it is implemented as web service, which offers flexibility in the programming languages.

**StreamingRec** (Jugovac et al. 2018) is written in Java and offers a variety of pre-built news recommendation algorithms for implementation and comparative evaluations. It simulates real-world news recommendation scenarios.

**CLEF NEWSREEL and Open Recommendation Platform (ORP),** CLEF NEWSREEL platform was designed to encourage researchers to develop novel recommenders using the Plista dataset and evaluate them in real time through ORP. ORP consists of distributed systems where recommendation providers and consumers interact over a standardized protocol to deliver recommendations. The researchers used CLEF NEWSREEL for online evaluation as well as the replay-based (simulation or offline) evaluation (Domann and Lommatzsch 2017; Kumar et al. 2017). It also includes the Idomaar framework, the Plista dataset and offers a few online algorithms and data analysis techniques.

Among these frameworks, Idomaar and Apache Mahout frameworks were developed for general recommender systems, whereas CLEF NEWSREEL, streamingRec and MIND were designed specifically for NRS. CLEF NEWSREEL is obsolete now. The MIND platform is still active.

# 4 Major Challenges in News Recommender Systems and Conventional Solutions

In this section, we discuss the major challenges of the NRS and their solutions. A few challenges such as cold-start, data sparsity have been reviewed in the previous survey (Karimi et al. 2018). They are common to the general recommenders too. So, we decide to skip them in this survey. We include two challenges (timeliness and user modeling) that have been discussed before, but we try to provide some new insights and perspectives in our discussion. We also identify the news content quality as an emerging challenge, which is not discussed before.

Here, we provide a categorization of the conventional solutions from the state-of-the-art to address these major challenges. We use the term "conventional" to refer to the non-

---

neural solutions, and we leave the discussion of the deep-learning-based solutions to Section 5.

## 4.1 Challenge 1: Timeliness

The earlier an event is reported, the more newsworthy it becomes. According to the working notes of CLEF NEWSREEL challenge (Brodt and Hopfgartner 2014), a well-formed recommendation must respond to a request within a given time frame (100ms). It requires faster, real-time processing and much more computations to make recommendations for a large number of news articles that are found in the news domain. Popularity, recency, freshness, trends, uniqueness, and low latency are the characteristics that should be factored into an NRS to give timely suggestions.

**Solutions:** Several conventional techniques used in general recommenders have been applied to address the challenge of timeliness in NRS. These models are discussed below.

### 4.1.1 Time-decay Models

Recommendation algorithms designed to give more weight to recent items with sensitivity to time are called time-decay models (Ding and Li 2005; Xia et al. 2010). The term 'time-decay' refers to the decline in terms of value of data over time. To be able to accommodate the time-decay effect of news items, it is important to build an effective short-term preference model that can predict the recent news items to readers.

A naïve and popular time-decay model is to use sliding/timing windows. A timing window in a time-decay model considers only recent news items or rating data, with older data being discarded or weighted less (De Francisci Morales et al. 2012). In the literature, there are various reports about the sizes and weights of timing windows. Some authors (Fortuna et al. 2015; Okura et al. 2017; Sottocornola et al. 2018) state that the timing windows should not be of fixed size (large or small) and should be adaptive. In general, a larger timing window leads to concept drifts (target variables change their values with time) (Muralidhar et al. 2015; Sottocornola et al. 2018) and a smaller one would not have sufficient data to build a short-term preference model (Sottocornola et al. 2018).

### 4.1.2 Graph-based Solutions

The second group of algorithms is graph-based that models the sequential reading process in an NRS. Graph-based recommendation models represent the relationship between users and items using links (weighted or unweighted). These models are also used to predict the next-news items by modeling the sequential dependencies over user-item interactions.

Some representative models include: (i) Context Trees that provide news recommendations to anonymous readers based on their news browsing patterns (Garcin et al. 2013; Maksai et al. 2015), (ii) Browse-Graphs to model sequential patterns from the readers' consumption histories (Trevisiol et al. 2014), and (iii) Markov decision process that models the sequential reading process in an NRS (Khattar et al. 2017). These traditional models are intuitive solutions to model the sequential dependencies among user-item interactions. However, due to an increasing number of states, these models may fail to capture the complex patterns from a large amount of data, as in the news domain.

### 4.1.3    Popularity-based Solutions

The third group of models in the NRS is the popularity-based models. They are based on the popularity of news items in terms of clickthrough rate, or social ties on social network sites. A traditional method of including popularity in an NRS is to simply count the total number of visits on news articles (Doychev et al. 2015). However, calculating popularity based on top-N articles is prone to amplification (popularity bias or temporal bias), which is caused by exclusively selecting top-N articles while overlooking the good (N+1)th candidate article (s). In this case, some good articles are unfairly penalized during the hard cut-off, despite the fact that the differences between these articles and the top-N recommendations are negligible. This issue can be mitigated if recommendations are generated probabilistically with feedback loops in which an article's likelihood of being chosen is proportional to its current popularity (count) (Prawesh and Padmanabhan 2012). News stories can also be ranked according to their popularity in the popular micro-blogging sites like Twitter (Jonnalagedda et al. 2016). In some NRS, the trends are also used to determine the popularity of news items (Chakraborty et al. 2019).

Although popularity-based models are easy to implement, it does not ensure that all popular news is credible and truly popular. According to a report by nbcnews[25], false news stories are more popular, and they are 70% more likely to be retweeted than true stories.

Overall, traditional timeliness models may be limited in their ability to address dynamic user behaviors in an NRS.

### 4.2    Challenge 2: User Modeling

Typically, users' preferences are modeled in two ways: explicit feedbacks and implicit feedbacks (Knijnenburg et al. 2012). Explicit feedback data is quantifiable, e.g., the rating of movies by users on Netflix or products on Amazon or news items on Flipboard. Often in an NRS, it happens that a user may read the whole news article but does not explicitly specify the rating. In this case, we consider implicit feedbacks that act as a proxy for a user's interest. Examples of implicit feedback data include clicks on links, browsing history, reading time spent and percentage (5%, 50% or 75%) of scrolling a news story.

In an NRS, we need to consider several aspects of user modelling, such as anonymous news readers, profiling information for registered users, passive news consumption, negative implicit feedbacks, and relevance of readers' intents.

**Solutions:** We review the pertinent literature to find out different user modeling techniques used in the NRS. These models are discussed below.

### 4.2.1    Stereotypical User Modeling

The first approach is stereotypical user modeling A stereotype is a collection of characteristics that frequently co-occur in people (Rich 1979). In this approach, a user is assigned to a class of users, and predictions about users' preferences are inferred from prior information about the class. When we do not have the complete background knowledge about a user, we can use this modelling technique. Well-known stereotypes in

---

the NRS are based on geolocation (Asikin and Wörndl 2014; Garrido et al. 2015; Robindro et al. 2017) and on users' habits (Constantinides and Dowell 2018).

Though stereotyping allows users to be classified into different groups, there are two issues with stereotyping in the NRS: (i) there is no way to learn a completely new stereotype, and (ii) too much stereotyping may result in segregated user groups or filter bubbles among like-minded users.

### 4.2.2 Feature-based User Modeling

The second approach is feature-based user modeling**.** A news article's content typically consists of features such as categories, headlines, sources, and topics. These features are extracted using statistical text representation methods such as bag-of-words (BoW), TFIDF, Hashing, and Word2vec. If the content of a news story is similar to the one that the user has previously read, it is recommended to the user. A general limitation of these traditional methods is that they do not consider the semantic (meaning in text) and contexts (situation in which a reader interacts with the news) while making news recommendations.

A user interest profile typically consists of long-term interests that can be captured from keywords extracted from a user's previous readings (Oh et al. 2014) or from his/her implicit feedback information (Muralidhar et al. 2015). Because users' preferences in a news domain are quite volatile and many users are anonymous, it is difficult to have complete profiling information using these statistical methods. These traditional methods are also limited to capture the time-ordered dependencies in readers' preferences.

### 4.2.3 Collaborative Filtering

User modeling based on users' interactions, i.e., the collaborative filtering approach, to make recommendations does not require analysis of item features. These methods collect interests from similar users and store them as histories. However, if the temporal distinction in user preferences is not preserved, an NRS may be unable to effectively predict the next news article based on similar user preferences. This necessitates that an NRS incorporate the time sequence of user behaviors into the traditional CF approach (Xiao et al. 2015; Khattar et al. 2017; Raza and Ding 2019).

### 4.2.4 Knowledge-based User Modeling

Knowledge-based user modeling approach is often used to apply semantics (Khattar et al. 2017), ontologies (Agarwal et al. 2013) or other contexts (situation in which a user is currently in) to model users' preferences (Wang et al. 2018b). In a few NRS, OWL ontologies based on IPTC[26] (International Press Telecommunication Council) standards (Agarwal et al. 2013) and free knowledge bases such as Wikipedia or Microsoft Satori (Wang et al. 2018a) are used to build rich content profiles. These models allow for the reuse of domain knowledge but creating a new knowledge base may be expensive.

### 4.2.5 Microblogging-based User Modeling

Microblogging user modelling makes use of social media platforms (such as Twitter) to model users' preferences and provide them with personalized and trending news services.

---

[26] https://iptc.org/

There are numerous examples in the literature where users' interest profiles were inferred from microblogs (De Francisci Morales et al. 2012; Gu et al. 2014; Jonnalagedda et al. 2016). Although microblogging provides rich user interaction data, additional measures are required to assess the quality of such content (Kang et al. 2015). For example, communications and discussions in microblogs in comparison to curated news stories are usually not much trustworthy (Kang et al. 2015; Cucchiarelli et al. 2018).

In general, traditional methods for user modelling in an NRS are not very successful. In an NRS, user modelling should include not only users' histories, but also their short-term, seasonal, diversified, and sequential interests.

### 4.3    Challenge 3: Quality Control of the News Content

With the majority of news media moving online, the initial difficulty for the research community was figuring out how to efficiently handle and evaluate the massive volume of unstructured information (most internet news is in textual format) in real time. Big data technology (e.g., Spark, Hadoop and cloud technology) has partially resolved the efficiency and scalability issue, while the latest development in the NLP field (e.g., the embedding-based and deep learning models) has partially resolved the feature engineering issue. The new and unsolved challenge is the quality control of the news content.

The researchers from social science usually do two types of content analyses in the news domain: quantitative and qualitative (Hamborg et al. 2019). To evaluate the quality of the news content, qualitative analysis usually requires the gold-standard test (human interpretation), which is a time-consuming task. Quantitative analysis determines the frequency of specific words or phrases in news articles, as well as other statistical features of news, such as the number of articles published on a news topic, the number of words per story, the placement of the news story on the website, and so on. In comparison to social science, quality control of the news domain is a new and understudied research topic in computer science.

The issues with the content quality in the limited research can be summarized as: duplication, lack of semantics, spamming and biases in the news items.

**Duplication**: Similar content appears at multiple locations (URLs) from different news sources (Doychev et al. 2015; Okura et al. 2017; Robindro et al. 2017). This can affect the ranking of news articles and is likely to bore the readers with repeated recommendations.

**Lack of semantics**: Multiple jargons and slangs with missing semantics can often be found in news stories (Mohallick and Özgöbek 2017). They are hard to interpret using available NLP libraries.

**Spamming**: Clickbaits (catchy news headlines) are used to trick news readers so that they click heavily on the news sites (Chakraborty et al. 2016). It is difficult to extract the hidden meanings from the clickbait that is used to manipulate readers. Even when these semantics are captured, the tactics of the used spamming techniques may later be modified.

**Biases:** The style in which the news stories are written and the tone in which they are presented reflect the biases of the publishers, authors and the media group (Kang et al. 2015). The hyper-partisan bias (bias from publishers) is a major issue in today's news.

**Solutions:** We reviewed the pertinent literature to find out how different authors addressed quality control issues in the NRS. These methods are discussed below.

### 4.3.1 Duplication Detection Methods

The traditional statistical methods such as TF-IDF or BoW techniques based on content features are used to recommend similar news articles to the target user (Doychev et al. 2015). But similar news articles are often repeated in the sense that they refer to the same news story presented in different ways from different publishers. Some duplication detection methods are discussed in the previous NRS research. For example, in one NRS (Okura et al. 2016), a threshold was used to filter out repetitive news articles (with similarity greater than a pre-defined maximum value). Another NRS (Robindro et al. 2017) addresses repetitive recommendations by clustering similar articles (using k-means) and then selecting a representative from each cluster. These traditional clustering-based methods are incapable of generating content embeddings for a large number of news articles and detecting duplication.

### 4.3.2 Semantics-based Methods

To improve the quality of news recommendations, a few authors addressed the lack of semantics in NRS. For example, semantics about news stories are captured from the news structure metadata (taxonomy) in one NRS (Ilievski and Roy 2013). While this method (Ilievski and Roy 2013) focuses on higher-level semantics, we find it falls short of providing a complete representation of semantics from news bodies, titles, and so on. In another NRS (Khattar et al. 2017), the ontology is used to introduce semantic similarity among news articles. Another NRS uses the concepts and named entities from Wikipedia pages to capture the semantics of news articles (Cucchiarelli et al. 2018). Other issues that are not addressed by these methods include changing ontologies, scalability, and multilingualism.

### 4.3.3 Bias Detection Methods

These methods can detect bias in news articles. Sentiment analysis techniques have been used in a few NRS to detect sentiment bearing words from news text (Ilievski and Roy 2013; Wang and Wu 2015; Khattar et al. 2017; Cucchiarelli et al. 2018). The exploration-exploitation principle is used in another NRS to reduce bias in news articles (Boutet et al. 2013). These bias detection methods are limited, and more research is needed to detect the level of bias and to mitigate with the new tactics of introducing biases in the data.

### 4.3.4 Clickbait Detection Methods

There is limited work in the NRS to address the clickbait (catchy deceptive headlines) problem. In one NRS, the clickbait can be distinguished from the regular news headlines through a classification method (Chakraborty et al. 2016). The method is trained on a clickbait dataset gathered from a few domains that publish a large number of clickbait articles: 'BuzzFeed', 'Upworthy', 'ViralNova', 'Scoopwhoop', and 'ViralStories'. The proposed classifier, then, identifies the clickbait headlines based on linguistic and syntactic nuances that appear more frequently in clickbait headlines.

Because the tactics of clickbait creators change over time, a typical classification model trained on a specific dataset from a specific time may suffer from data and concept drifts. As a result, to keep up with changing tactics, a classification model may need to be trained on a regular basis. Furthermore, the semantics and hidden patterns from the clickbait data should also be included in the classification models.

These papers and conventional solutions are summarized in Table 3. As can be seen in Table 3, the challenge that is addressed the most in NRS is user modeling, followed by the timeliness. The work on content quality is marginal and it needs more attention in NRS.

**Table 3.** Algorithms (alg.), challenge, solution, dataset, evaluation (eval) metric: accuracy (acc): beyond-accuracy (beyond-acc), and evaluation protocol (protocol) for NRS papers

| Paper | Alg. | Challenge | Solution | Dataset | Eval. Metric | Protocol |
|---|---|---|---|---|---|---|
| (Park et al. 2009) | CBF | Content quality | Bias detection | Private - Google news | F1-score (acc) | Offline, User Study |
| (Xia et al. 2010) | CF | Timeliness | Time-decay | Private - Alibaba.com | CSI (beyond-acc), prec (acc) | Offline |
| (De Francisci Morales et al. 2012) | CBF | Timelines User modeling | Time-decay Microblog | Private - Twitter, Yahoo | MRR (acc), cov (beyond-acc) | Offline |
| (Agarwal et al. 2013) | CBF | User modeling | Ontology | Private - RSS news feed | prec, rec (acc) | Offline |
| (Prawesh and Padmanabhan 2012) | CBF | Timeliness | Popularity | - | - | - |
| (Boutet et al. 2013) | CF | User modeling Content quality | Feature-based Bias detection | Private - Arxiv,Digg[27] | prec, rec (acc) | Offline, User Study |
| (Ilievski and Roy 2013) | CBF | User modeling | Knowledge-based | Private - German news papers | Rank (acc) | Offline |
| (Li and Li 2013) | CBF, CF | User modeling | Feature-based | Private - not mentioned | prec, rec, F1-score, NDCG (acc), div (beyond-acc) | Offline |
| (Jonnalagedda and Gauch 2013) | CBF, CF | Timeliness | Popularity-based | Private - Twitter, CNN and BBC | Accuracy (acc) | Offline, User Study |
| (Garcin et al. 2013) | CBF | Timeliness | Graph-based | Private - Tribune de Geneve and 24heures.ch | Personalized (acc), nov (beyond-acc) | Offline |
| (Gu et al. 2014) | CBF | User modeling | Microblog | Private - news.sina.com. | F1, prec, rec (acc), div, nov (beyond-acc) | Offline |
| (Asikin and Wörndl 2014) | CBF | User modeling | Stereotypical | Private | seren (beyond-acc) | User study |
| (Trevisiol et al. 2014) | CBF | Timeliness | Graph-based | Private - Facebook, Twitter, Reddit | prec, MRR (acc) | Offline |
| (Oh et al. 2014) | CBF | User modeling | Feature-based | Private - Korean news | HR (acc) | Offline |

| (Muralidhar et al. 2015) | CF, CBF | Timeliness User modeling | Time-decay Feature-based | Private - Washington Post | HR (acc) | Offline |
|---|---|---|---|---|---|---|
| (Xiao et al. 2015) | CF | Timeliness, User modeling | Sequential Collaborative | Private Chinese news | prec, rec, F1 (acc) | Offline |
| (Maksai et al. 2015) | CBF | Timeliness | Popularity | Private - Swissinfo.ch, Yahoo FrontPage, lePoint.fr | div, nov, seren, cov (beyond-acc), RMSE, CTR (acc) | Online, Offline |
| (Jenders et al. 2015) | CBF | Content quality | Feature-based | Private - New York Times | seren (beyond-acc) | User Study |
| (Garrido et al. 2015) | CBF | User modeling | Stereotypical | Private - heraldo.es | - | Offline |
| (Lu et al. 2015) | CBF, CF | User modeling | Feature-based, Collaborative | Private - Bing Now news | Accuracy | Offline |
| (Doychev et al. 2015) | CBF, CF | Timeliness | Popularity | Plista | prec, CTR (acc) | Online, Offline |
| (Ma et al. 2016) | CBF, CF | User modeling | Stereotypical, Collaborative | Private - Bing Now news | MAP, MRR, CTR (acc) | Offline |
| (Jonnalagedda et al. 2016) | CBF | User modeling | Microblog | Private - Twitter | NDCG (acc) | Offline |
| (Chakraborty et al. 2017) | CBF | User modeling | Clickbait removal | Private - BuzzFeed & Wikinews | prec, rec, F1 (acc) | Offline, Online |
| (Viana and Soares 2016) | CBF, CF | User modeling | Feature-based | - | - | User Study, Simulation |
| (Rizos et al. 2016) | CBF | User Modeling | Feature-based, Microblog | Private - Reddit | Accuracy | Offline |
| (Okura et al. 2016) | CBF | Content quality | Duplicate detection | Private - Yahoo FrontPage | CTR (acc) | Online |
| (Guan et al. 2017) | CBF, CF | Timeliness | Deep neural network | Private - news.sohu.com | prec, rec, F1 (acc) | Offline |
| (Robindro et al. 2017) | CBF | User modeling | Stereotypical | BBC news | NDCG (acc) | Offline |
| (Khattar et al. 2017) | CF | Timeliness User modeling | Session-based, Knowledge-based | Private - Veooz.com news | MAP, Hit rate, NDCG (acc) | Offline |
| (Okura et al. 2017) | CBF | Content quality Timeliness | Duplicate detection, Time decay | Private - Yahoo FrontPage | AUC, MRR, NDCG, CTR (Online) (acc) | Online, Offline |
| (Kumar et al. 2017) | CBF | Content quality | Knowledge-based | Plista | HR, NDCG (acc) | Online, Offline |
| (Cucchiarelli et al. 2018) | CBF | User modeling | Microblogging | Private - Twitter, Wikipedia | Saliency, seren (beyond-acc), MAP (acc) | Online, User Study |
| (Constantinides and Dowell 2018) | CBF, CF | User modeling | Stereotyping | Private - Habito.com News | Accuracy | Offline, User Study |
| (Jugovac et al. 2018) | CF | User modeling Timeliness | Feature-based Session-based | Outbrain.com, Plista | MRR, F1-Score (acc) | Online, Offline |
| (de Souza Pereira Moreira 2018) | CBF CF | Timeliness | Session-based | Adressa, Globo.com | Rec, NDCG (acc) | Offline |
| (Sottocornola et al. 2018) | CBF CF | Timeliness | Time-decay, Session-based | Private - not mentioned | prec (acc) | Offline |
| (Lin et al. 2012) | CBF | User Modeling | PMF | Private – not mentioned | Accuracy | Offline |
| (Yan et al. 2012) | CBF | User Modeling | NMF | Tweet data. News data crawled from sogou.com | Accuracy | Offline |
| (Xia et al. 2014) | CF | Content Quality | BPR | Private - not mentioned | Accuracy | Offline |
| (Wang et al. 2015) | CBF | User Modeling | TF | Not mentioned | - | - |
| (Shu et al. 2019) | CBF | User Modeling | NMF | BuzzFeed, PolitiFact | Accuracy | Offline |
| (Gharahighehi and Vens 2019) | CBF | User Modeling | BPR | Private- not mentioned | Accuracy | Offline |
| (Raza and Ding 2019) | CBF | User Modeling, Timeliness | MF (baseline predictors) | NYTimes - crawled | prec,rec (acc) | Offline |
| (Raza and Ding 2020a) | CBF | User Modeling | GLM with CF | NYTimes - crawled | prec, rec (acc), div, nov (beyond-acc) | Offline |

## 5 Deep Learning Models for News Recommender Systems

In this section, we cover Deep Learning (DL)-based solutions that have been widely applied in the NRS research in recent years. Many of the challenges that an NRS faces are seen to be addressed using these DL models. These methods build user models in different ways than the traditional recommendation models, and they deal with timeliness and other NRS-related issues in a more advanced way. There are certain advantages of DL that makes it a preferable approach in the NRS over some conventional solutions, which are discussed below.

The first advantage of DL is its strength when dealing with the content-based recommendation. It is inevitable for a typical CBF method to handle massive amount of data that is also multimodal (text/audio/video). For instance, when dealing with the textual data (news story, reviews, comments, tweets, etc.), images or videos, the deep neural methods like CNNs/RNNs (An et al. 2019) or language models like BERT (Devlin et al. 2018) are indispensable in the representation learning (feature learning) tasks.

The second significant advantage of DL is its ability to learn multiple interactions between the users and items. The DL-based NRS (de Souza Pereira Moreira 2018) also demonstrates sufficient performance gains over traditional CF methods (Xiao et al. 2015) in learning rich user-item interactions from the news data.

The third strength of DL is in sequential modeling. The sequential modeling task is an important approach for mining the temporal dynamics (changing user behavior over time) and session-based news recommendation tasks. Compared to this, the traditional CBF or CF are often built on the static datasets, where there is no consideration of the temporal or sequential factors.

The fourth strength of DL methods is in dealing with the cold-start and data sparsity issues of conventional recommendation methods. The cold-start and data sparsity problem in conventional NRS is the result of insufficient rating information. The DL can extract useful features from the news and user data, which improves the estimation of user and item profiles and, as a result, improves the recommendation accuracy.

Next, we discuss the DL-based models for news recommendations.

### 5.1 Multi-Layer Perceptron (MLP)

MLP is a feed-forward neural network in which there are multiple hidden layers between the input and output layer. In a recommender system, the MLP can be used to add non-linear transformations on top of a typical MF, to learn rich user-item interactions. For example, the NCF (He et al. 2017) uses the non-linearity of MLP to learn the user-item interactions in the CF model. The MLP is also used in a few NRS (Song et al. 2016; Yu et al. 2018) to learn useful representations from data.

Overall, MLP is a simple and efficient model that is used to create neural extensions of MF based models.

### 5.2 Autoencoder (AE)

AE is a neural network that learns to copy its input to the output in an unsupervised manner. It has an internal (hidden) layer that describes a code to represent input, and it is made up of two primary components: an encoder to map the input into code, and a decoder to map the code to reconstruct input. In a recommender system, the AE and its variants are often used to learn hidden patterns to reconstruct users' ratings from their historical interactions (Wu et al. 2016). AE methods are also used to compress a dataset into a lower-dimensional feature subspace while preserving most of the relevant information.

Denoising auto-encoders (a type of AE) are used in an NRS to create news article representations (Okura et al. 2017). In another NRS (Cao et al. 2017), the stacked AE are used to extract the low dimensional features from a sparse rating matrix.

Overall, the AE are effective at learning useful representations from news data (news content and user feedback) in a low-dimensional space.

### 5.3 Convolutional Neural Network (CNN)

CNN is a feed-forward neural network with convolutional layers and pooling operations and have achieved great success in the field of computer vision, particularly for automated diagnosis in medicine (Göçeri 2020a, b). A CNN typically has two sets of layers: (i) convolution layers for generating local features from the data; and (ii) pooling (or sub-sampling) layers to select only representative local features (i.e., features with the highest score via activation functions) from the previous (convolution) layer. Compared to MLP networks, the CNNs have fewer parameters and perform faster (He et al. 2018).

A CNN can extract useful features from the news data by using convolution operations (also known as kernels or filters) at varying levels of granularity, thus eliminate the need for manual feature engineering (Yu et al. 2018). They are frequently used to extract local text features from news headlines (Wang et al. 2018a; An et al. 2019; Wu et al. 2019a) or from entire news bodies (Zhu et al. 2019). The knowledge gained from these news representations is then used to make recommendations by computing the similarity between the candidate and the clicked news (Wang et al. 2018a; Zhu et al. 2019).

Overall, CNNs are useful methods for representing multimodal (text, audio, video) features from the news data.

### 5.4 Recurrent Neural Network (RNN)

RNN models are used to model variable-length sequence data. In a recommender system, the RNNs are often used to model sequential dependencies in the rating data and for session-based recommendation tasks (Hidasi et al. 2016). Two well-known variants of RNNs are Long short-term memory (LSTM) and Gated recurrent unit (GRU). The key difference between the two is that GRU does not need memory units as in LSTM, so, GRUs are faster to train. However, it is easier to learn longer sequences with LSTM.

The GRUs have been used in a few NRS to learn short-term users' preferences from the interaction histories (Okura et al. 2017; An et al. 2019; Zhang et al. 2019). The results demonstrated a significant improvement over the traditional temporal models, with slightly better performance of GRUs over LSTM (Okura et al. 2017).

Song et al. (Song et al. 2016) propose to learn user's short-term preferences using unidirectional LSTM (Song et al. 2016). The unidirectional LSTM only preserves information of the past. The improvement over unidirectional LSTM is made by replacing it with the bidirectional LSTM in another NRS (Kumar et al. 2017). Bidirectional LSTM runs the user input sequence in two ways, one from the past to the future (forward pass) and one from the future to the past (backward pass).

In a few recent NRS (de Souza Pereira Moreira 2018; An et al. 2019; Wu et al.), the GRUs are successfully used to learn users' short-term preferences. Some NRS (Zhu et al. 2019) also use LSTM networks for spotting users' preferences in shorter time periods. One NRS (Wu et al. 2019a) also adds the neural attention (Vaswani et al. 2017) on each state of RNN to get rich sequential features during different clicking time.

Overall, RNNs are useful for modeling session-based and sequence-based recommendation tasks. These models can also be used to incorporate additional news-related information during different temporal steps (An et al. 2019).

### 5.5 Neural Attention

The neural attention (Vaswani et al. 2017) is based on the idea that a model pays attention to a specific part when processing a huge amount of information. Neural attention has gained remarkable success in a variety of machine learning applications, including language modelling, image captioning, and text classification. The attention mechanism is also employed in recommender systems to filter out noisy content and to select the most representative items.

In some NRS (Wang et al. 2018a; An et al. 2019; Wu et al. 2019a), the attention is used at the word-level to learn informative words from the news content. The attention is also applied at the news-level to model the informativeness of different kinds of news information for learning useful news representations (Wu et al. 2019b). For example, if news headline is more important than the other pieces of news (news body, topic, taxonomy), then it should be weighed more. Because the informativeness of the same words and news may differ amongst users, the idea of personalized attention network is applied in another NRS (Wu et al. 2019a). The personalized attention network uses the embedding of the user information as the query vector of word and news-level attention networks; and attends to significant words and news in different ways based on user preferences.

The attention mechanism are useful to learn news and user representations from the neural networks and is the backbone of Transformer models (Devlin et al. 2018).

### 5.6    Graph Neural Network (GNN)

Recently, the GNN models (Scarselli et al. 2008) have gained increasing popularity in a variety of domains, including social networking, recommendation systems, search engines and related. GNN is a type of neural network that operates directly on the structure of the graph. Essentially, each node in the graph is linked to a label, and the task is to predict the label. GNNs are used for classification tasks, such as text classification, labelling sequences, machine translation, and for prediction tasks.

GNNs are used for the recommendation tasks in a few recent NRS (Wu et al. 2019c; Lee et al. 2020; Sheu and Li 2020; Yang et al. 2020; Ge et al. 2020). GNewsRec (Ge et al. 2020) is a GNN-based news recommendation system that constructs a reader-news-topic graph to learn the embeddings from the news features and reader's clicks. Both representations (news and reader) are then used to determine the click probability of the candidate news to be recommended next.

Overall, the GNNs are promising models that yield outstanding outcomes when paired with the attention mechanism (Wu et al. 2020c).

### 5.7    Transformers

Transformer model, introduced in the neural attention paper (Vaswani et al. 2017), has achieved state-of-the-art performance in NLP tasks. Transformers are intended for handling the sequential data in the same manner as RNNs. However, in contrast to RNNs, the transformers do not require the sequential data to be processed in order (one after other). Instead, the transformers process sequential data in parallel. The main crux of a Transformer is the self-attention layer. The self-attention looks at an input sequence and decides at each step which other parts of the sequence are important.

The idea of Transformers is taken from the transfer learning, where a big language model is usually trained on billions of words, and the knowledge from the big model is transferred to similar smaller NLP tasks. For example, the Google BERT (Devlin et al. 2018) model is pre-trained on a large corpus of unlabeled text including the entire Wikipedia and Toronto Book Corpus, and is used to train other models on downstream NLP tasks, for the purpose of making better predictions. Well-known Transformer models are BERT, BART, ALBERT, GPT-2, RoBERTa, and other listed here[28].

The deep bidirectional self-attention BERT is employed to model the sequences in user's behavior for the click prediction task in a recommender system (Sun et al. 2019). A recent NRS (Wu et al. 2021) is built on the same idea to use the BERT for the task of news recommendations.

---

[28] https://huggingface.co/transformers/

### 5.8 Reinforcement Learning (RL)

Deep RL methods are based on trial-and-error paradigm and have demonstrated human-level performance across various domains such as games, robotics, finances and even recommenders (Francois-Lavet et al. 2018). RL consists of five components (agents, environments, states, actions, and rewards) to get knowledge from the raw data. Deep Q-Learning (DQN) is a RL strategy that, given a current state, helps to find the maximum expected future reward of an action. The DQN structure has been applied in an NRS (Zheng et al. 2018) to model the dynamics in users' preferences and that of news content. The RL models can also be used to define the best sequence of decisions through the interaction with the news environment and observation of rewards (clicks).

### 5.9 Summary

The DL methods have proven to be very successful in building the NRS and appear to have the great potential to be further used in the future. Despite the success of these methods, one limitation is noted. It is that the current NRS research (including DL-based models) focus too much on the accuracy of the models to provide recommendations to the users. The aspects beyond accuracy such as novelty, serendipity, diversity and a composite user model are not very much covered in these approaches. These deep learning solutions and the challenges they have addressed for the NRS are summarized in Table 4.

**Table 4:** Algorithms (alg.), DL mechanism, dataset, evaluation (eval) metric: accuracy (acc): beyond-accuracy (beyond-acc), and evaluation protocol (protocol) for NRS papers

| Paper | Alg. | DL mechanism | Dataset | Eval. Metric | Protocol |
|---|---|---|---|---|---|
| (Song et al. 2016) | CBF | RNNs (LSTM) for timeliness in user modeling for short-term preferences. Sent2Vec (BoW) for news representation. | Private – crawled, not mentioned | prec, rec, F1, AUC, MAP, MRR (acc)y | Offline |
| (Okura et al. 2017) | CBF | Denoising AE for news representations and RNNs (LSTM) for timeliness in user modeling for short-term preferences. | Private -Yahoo Japan news | AUC, MRR, DCG, CTR (acc) | Offline |
| (Wang et al. 2017) | CBF + CF | CNNs for news representations; Attention for user modeling | Private - crawled | AUC, prec, rec, F1 (acc) | Offline, online |
| (Kumar et al. 2017) | CF | Doc2Vec for news representation, RNNs (Bidirectional LSTM) for timeliness in user modeling for short-term preferences. | Private - crawled | HR, NDCG (acc) | Offline |
| (Lian et al. 2018) | CBF | CNNs for news representations, Attention for user modeling | Private – Bing | AUC, Logloss (acc) | Offline |
| (Wang et al. 2018a) | CBF | Knowledge graph with CNNs for news representations, Attention for user modeling. | Private - Bing news logs | CTR (acc) | Offline |
| (Zheng et al. 2018) | CBF | RL, Deep Q-Network for news and user representations | Private – not given | CTR, prec, NDCG, (acc), div (beyond-acc) | Offline |
| (Yu et al. 2018) | CBF | CNNs for news text, AE for images representation, MLP to recommend. | Private– frnews.ifeng.com.ca | Accuracy | Offline |

| (de Souza Pereira Moreira 2018) | CBF, CF | Word2Vec+ CNN for news representations, RNNs (GRUs) for timeliness in user modeling for short-term preferences. | Globo.com, Plista, Adressa | HR, MRR, Accuracy (acc); cov, nov, div (beyond-acc) | Online and Offline |
|---|---|---|---|---|---|
| (Zhang et al. 2019) | CBF | CNNs for news representations, RNNs (GRUs) for timeliness in user modeling for short-term preferences. | Adressa | prec, rec, AUC (acc) | Offline |
| (Zhu et al. 2019) | CBF | CNNs for news representations, RNNs (GRUs) for timeliness in user modeling for short-term preferences. | Adressa | prec, rec, AUC (acc) | Offline |
| (Cao et al. 2017) | CBF | Stacked AE | Movielens (not news) | prec, rec, AUC (acc) | Offline |
| (An et al. 2019) | CBF, CF | CNNs for news representations, RNNs (GRUs) for timeliness in user modeling for short-term preferences. | MIND | AUC, NDCG, MRR (acc) | Offline |
| (Wu et al. 2019a) | CBF | CNNs for news representations, Attention for user modeling | MIND | AUC, NDCG, MRR (acc) | Offline |
| (Wu et al. 2019b) | CBF, CF | CNNs for news representations, Attention for user modeling | MIND | AUC, NDCG, MRR (acc) | Offline |
| (Wu et al. 2019c), (Ge et al. 2020) | CBF, CF | Transformer for news representations, GNNs for user modeling. | MIND | AUC, NDCG, MRR (acc) | Offline |
| (Wu et al. 2020a) | CBF, CF | Attention for news and user representations, MSE loss in news module for diversity | MIND | AUC, NDCG, MRR (acc), Senti (beyond-acc) | Offline |
| (Lee et al. 2020) | CBF, CF | GNNs for news and user representations | - | - | - |
| (Yang et al. 2020) | CBF, CF | Knowledge Graphs+ CNNs for news; Attention for user modeling. | Private-Wikidata, Weibo | prec, rec, F1, AUC (acc) | Offline |
| (Wu et al. 2021) | CBF, CF | BERT for news representations; Attention for user representation. | MIND | AUC, NDCG, MRR (acc), Senti(beyond-acc) | Offline |

As can be seen in Table 4, the user modeling is the most widely addressed challenge in the DL-based NRS. The challenge of timeliness is also addressed in these models. Usually, the sessions-based recommendation tasks are used to model timeliness in users' short-term preferences. These sessions are created in a chronological order of item click events or the publication time of the news items. There is not much work seen in terms of addressing the content quality challenge in these methods. Among all the DL methods, the CNNs and RNNs are popular choices for article and user representations. The attention mechanism can be found in the most recent DL papers. GNNs (with addition of attention) and the Transformers (primarily based on the neural attention) are also used in some recent papers. The accuracy metric and offline protocol remain the popular evaluation methods used in the DL-based NRS. There are also other useful DL models that are not seen in the NRS work lately, which we will discuss in Section 7 (Discussion).

# 6 Effects of News Recommendation Algorithms on Readers' Behavior

News organizations such as BBC, New York Times, The Guardian, and such, have worked hard to provide more personalized news stories to readers via their websites and applications. These recommendations are tailored to readers' preferences based on the topics of interest they have indicated in their profiles or, in some cases, the content they have recently consumed. It is a great accomplishment to provide readers with everything that truly reflects their interest. However, relying solely on machine learning algorithms, as in recommender systems, is not without risk. They are thought to have a negative impact on the production of news (fake news, exaggerated news, racism, persecution, stereotypes, and so on), readers' psychology, consumption behaviors, and overall user experiences with an NRS.

Although these negative impacts are being recognized in the literature of computer science, there is only a limited amount of work (Nguyen et al. 2014; Allcott and Gentzkow 2017; Möller et al. 2018; Helberger 2019) that briefly touch the issue (post-algorithmic effects of news recommendations on readers' behavior). This issue has been widely discussed in other disciplines such as information science and mass communication, where they blame news recommendation algorithm developers for making poor design choices.

The birth of social media, fake news, and polarized political media groups has been blamed for the effects of news recommendations on user behaviors (Allcott and Gentzkow 2017). Some authors (Beam 2014; Quattrociocchi et al. 2016; Anspach 2017) see social media interference in news media as a threat to democracy. For example, Quattrociocchi et al. (Quattrociocchi et al. 2016) conduct a study on the Facebook group user engagement data to determine whether echo chambers exist on social media. According to their findings, social network users create like-minded echo chambers on certain issues, limiting their exposure to counter-attitudinal behavior.

According to our findings, this topic has a high social relevance in a variety of disciplines, including computer science, journalism, political science, and economics. We gathered some statistics from Pew Research Center[29] reports, which conducted extensive surveys on these issues. Following these steps, we identify the major effects on users' behaviors. We also discuss the possible mitigation strategies in this section.

## 6.1 Post-Algorithmic News Recommendation Effects

**Filter bubble** corresponds to intellectual isolation caused by personalized searches or algorithms to selectively assume the information an individual wants to see (Pariser 2011).

**Echo chamber** refers to an information bubble around a user, where the user is only exposed to articles that reinforce their existing beliefs (Flaxman et al. 2016).

**Polarization** refers to the divergent views on policy (politics, religion, beliefs) into ideological extremes (Dandekar et al. 2013). The frequent interactions between like-minded persons result in polarization.

---

[29] http://www.pewresearch.org/

**Fragmentation of the public sphere** refers to the disintegration of the shared public sphere into smaller publics where the citizens in those spheres become are less aware of outside issues (Helberger 2019).

**Dehumanization** refers to the control of human judgement through predictive modelling without readers knowing how it is done. All human decisions are overtaken by artificially generated logic (Page et al. 2018).

**Biased assimilation** refers to biases in readers caused by algorithms. Users begin to process new information in a biased manner, which eventually reinforces them to critically examine disconfirming evidences (Dandekar et al. 2013).

**Denial to Counter-attitudinal behavior,** Counter-attitudinal behavior is defined as behavior that does not align with one's points of view but is valued and regarded as a high level of exposure to different points of view. (Beam 2014). Denial to counter-attitudinal behavior is an issue caused by filter bubble or echo chambers.

**Reinforced digital gate-keeping** refers to the selection and extraction of all news through digital gates (recommenders) with no human judgement (Möller et al. 2018).

**Deep Fakes** refers to media created by artificial neural networks that takes a person in an existing image/ video and replace with someone else, for example[30], the deep fake of Obama public announcement, and Donald Trump speaking informally. Deepfakes are created using social media, which has resulted in fake news and other conspiracy theories.

## 6.2 Mitigating Effects of News Recommendations on User Behavior

We have reviewed the state-of-the-art solutions to mitigate the effects of news algorithms on readers' behavior. First, we discuss the solutions from the state-of-the-art NRS papers and then we discuss other solutions in Section 7.

### 6.2.1 Selective Exposure

Selective exposure research is taken from Festinger's cognitive dissonance theory (Festinger 1962), a discipline of psychology that states people prefer to view information that supports their own perspectives (Hart et al. 2009). According to this theory, dissonant information (information that does not match with the user attitude) increases uncertainty and discomfort in a user. As a result, the user may read information that is pro-attitudinal (congruent with user behavior) and try to avoid information that is counter-attitudinal (conflicting with his perspectives). However, empirical research (Brundidge 2010) in selective exposure indicates that readers may also want to select and read different news stories to gain knowledge for both pro- and counter-attitudinal information.

Garret (Garrett 2009), for example, demonstrates through a user study that during the elections days, people tend to search online news about their favorite candidates. Further, the same participants also went on to search online news for the opposing candidates and read about their perspectives. This finding contradicts the pro-attitudinal user behavior.

Beam (Beam 2014) demonstrates through a user study that during selective exposure, users only select news stories that match with their own preferences. While doing so, they may be presented with news stories that contradict with their own beliefs; in this case, they may still want to read them so that they can form their own opinion on a particular issue. Flaxman et al. (Flaxman et al. 2016) support this, demonstrating through a large-scale user study that selective exposure during online news consumption exposes readers to information that does not always align with their political beliefs.

---

[30] https://www.creativebloq.com/features/deepfake-examples

Another group of researchers (Flaxman et al. 2016; Newman et al. 2018) believe that social media users are much more likely to encounter sources they would not normally encounter, exposing them to opposing viewpoints. Flaxman et al. (Flaxman et al. 2016) conduct a user study and analyze the web browsing histories of 50,000 US citizens who regularly read online news. The results demonstrate that the usage of social networks and search engines exposes users to counter-attitudinal information. The Reuters Report 2018 (Newman et al. 2018) also presents a user study and affirms the previous research that social media plays a role in increasing users' exposure to news.

Dandekar et al. (Dandekar et al. 2013) use DeGroot's graphical model of opinion to address the polarization problem in the news domain, in which individuals update their opinions based on a weighted averaging of their current opinions and those of their neighbors. Herlberger (Helberger 2019) also proposes a democratic recommender system providing news readers with a diverse mix of news recommendations.

Overall, more research is needed to include selective exposure in the design of an NRS.

### 6.2.2    Diversity-Aware Algorithms

These algorithms take diversity into account at various stages of the recommendation process, such as during the re-ranking process (after recommendations are generated) or the optimization phase (the recommendation process). A recommendation algorithm is typically programmed to promote exposure to unpopular items (long tail items) during the re-ranking process. During the optimization phase, the recommendation algorithm is tailored so that diversity, as well as the (built-in) accuracy objective, are included in the recommendation process. The news topics, writing styles, tags, perspectives, contexts, and ideologies are some of the factors that are considered to be diversified in an NRS (Resnick et al. 2013; DiFranzo and Gloria-Garcia 2017; Möller et al. 2018; Helberger 2019). Möller et al. (Möller et al. 2018) also propose to incorporate the diversity in an NRS as a democratic function identifiable in news articles, subjects, tones, styles of writing and political content.

In an earlier NRS (Rao et al. 2013), the news recommendation list is expanded by using the news taxonomy information to find relevant news items from encyclopedia websites. In another NRS (Zheng et al. 2018), the multi-arm bandit methods with exploration-exploitation optimization is used to tradeoff between accuracy and diversity. In a recent NRS (Raza and Ding 2020a),  the diversity is included through the use of regularization (Ridge regression for accuracy and Lasso regression for diversity) during the optimization phase. The diversity is, then, balanced with high accuracy in the model.

**Aspect:** An aspect is a collection of attributes, components, or services that can be used to categorize information. The aspects can diversify the news recommendations by providing readers with different perspectives on a news topic. In one NRS (Park et al. 2009), the news events are classified based on various aspects (topics) and then users are provided with different perspectives on news. Although there has been little work in aspect-level presentation in an NRS, it can be very useful to classify or cluster news articles based on other aspects (styles, tags, categories, sentiments) to make recommendations.

### 6.2.3   Nudge theory

This refers to giving subtle nudge (touch or push) in the form of small design changes that encourage users to make other choices in their general interest (van der Heijden and Kosters 2015). Nudging is a behavior change strategy that motivates people to achieve goals, and it can influence the behavior of news readers.

There have been some cases in real-world, where algorithms have been manipulated to steer readers toward fake news. For example, YouTube was manipulated consistently alongside the Guardian news to nudge readers towards sensational and fake news during US elections 2016 (DiFranzo and Gloria-Garcia 2017). Recently, the news websites are used in conjunction with the social media add-ons to spread the anti-vaccination misinformation and the rumor that incorrectly compared the number of registered voters in 2018 to the number of votes cast in US Elections 2020[31]. The implications of such news are seen in the anti-vaccine movements preventing the global fight against the COVID-19, or in the post-election unrest.

Despite these negative examples, nudges can be extremely beneficial when used transparently and ethically. Algorithms can be programmed to guide users toward more politically balanced news consumption and exposure. Resnick et al. (Resnick et al. 2013) design an interface (a browser add-on) that nudge the users to select more news rather than just relying on the algorithmic recommendations. There is also some work that demonstrates the general design and architecture of a smart nudge in a recommender system (Karlsen and Andersen 2019). Algotransparency[32] is also an information group that informs citizens on how people are nudged from an initially neutral search on YouTube to the progressively biased information throughout each subsequent phase of the recommendation cycle.

Overall, nudges, if used correctly, can help users make wise choices through selective exposure. However, it is difficult to observe users' behaviors excessively during nudging.

### 6.2.4   Trade-Off among Various Evaluation Measures

Maksai et al. (Maksai et al. 2015) quantify the trade-off between different metrics such as accuracy-coverage, accuracy-diversity, accuracy-serendipity, diversity-serendipity to test the performance of their recommendation algorithms. The results show that accuracy, when combined with beyond-accuracy measures, improves user behavior within an NRS.

Concerns about the potential negative consequences of personalization in the NRS have grown in recent years (Haim et al. 2018). Personalization is often the result of recommendations that align highly with users' preferences. Usually, a high accuracy results in higher personalization in a recommender system. However, we believe that personalization should not be completely ignored; otherwise, users may lose interest in an NRS where everything that is recommended is different or diverse. In fact, as demonstrated in a recent study, personalization can be balanced with reasonable diversity in an NRS (Raza and Ding 2020a).

---

[31] https://archive.is/OXJ60

[32] https://algotransparency.org/

Chakraborty et al. (Chakraborty et al. 2019) also take a closer look to balance three metrics: recency, importance (or popularity) and diversity in an NRS. In that, they propose a future-impact metric that takes the popularity signals from crowd-sourced information and the personalized information from the past news data to predict the impact of news stories for a news reader.

Overall, there has been little research in NRS that balances the built-in accuracy aspect with various aspects of quality evaluation and beyond-accuracy aspects.

**Table 5:** Post-algorithmic challenges in the NRS and the solutions

| Paper | Challenge | Algorithmic Cause | Effect on readers' behavior | Solution |
|---|---|---|---|---|
| (Dandekar et al. 2013) | Polarization | Narrowed readers' exposure | Causes denial to others' viewpoint | Degroot's graphical model of opinion |
| (Resnick et al. 2013) | Filter bubble | Over personalization | Creates filter bubbles, polarization | Selective exposure, diversity, nudge theory |
| (Beam 2014) | Counter-attitudinal | Over personalization | Affects readers' acceptance to opposing viewpoints | Selective exposure |
| (Li et al. 2014) | Filter bubble | Over personalization | Readers get bored of old news | Budgeted maximum coverage |
| (Maksai et al. 2015) | Filter bubble | Over personalization | Readers get bored of similar news stories | Trade-off among various evaluations |
| (Flaxman et al. 2016) | Filter bubble, Echo chambers | Ideological segregation | Affects voters and functioning democracies | Selective exposure |
| (Allcott and Gentzkow 2017) | Filter bubble, Echo chambers | Algorithmic fake news | Separates readers from contradictory perspectives | Selective exposure |
| (DiFranzo and Gloria-Garcia 2017) | Filter bubble, Echo chambers | Social media spread fake news | Causes readers' denial to others' opinions | Diversity-aware methods and nudges |
| (Möller et al. 2018) | Filter bubble | Accuracy-centric algorithms | Extreme opinions, misinterpreted facts | Diversity in re-ranking |
| (Helberger 2019) | Filter bubble | Over personalization | Affects negatively on the democracy | Exposure diversity, ban manipulative practices |
| (Zheng et al. 2018) | Filter bubble | Over personalization | Readers get bored of similar news stories | Dueling Bandit Gradient Descent for diversity |
| (Chakraborty et al. 2019) | Filter bubble | Over personalization | Readers tend to get bored of similar news stories | Future-impact metric |

***Summary:*** In the NRS, there is only a small amount of work that considers these factors (such as diversity, selective exposure, nudges, and aspects) in its design. The absence of such methods results in news recommendations that are entirely driven by the algorithmic logic of recommendation models or by the motivations of stakeholders (political figures, trading factors, etc.). The limited work addressing these issues is summarized in Table 5. There are also some suggestions that we discuss further in Section 7.

# 7    Discussion on Research Implications and Future Work

In this section, we highlight our major findings in this survey and discuss the research implications and future work.

### 7.1 Algorithmic Solutions and Major Challenges in NRS

Our review of selected publications revealed that NRS research is gradually gaining attention over time. One reason for this increase is the high conversion rate of traditional news media users to online news readers. This growth has provided researchers with numerous research opportunities to develop solutions to the unique challenges of news domain. Due to the rapid advancement of various DL methods, there has been a recent evolution in NRS research.

As discussed in Section 4, the traditional recommendation algorithms are not enough for building an NRS and can only partially address the challenges in the NRS. It requires a lot of modifications, extensions and variations on the standard recommendation approaches to meet the needs of news readers. The latent factor models as discussed in Section 3.2 and the DL models in Section 5, are two major classes of successful models to address the challenges faced by the NRS. The DL models, in particular, continue to be used in the recent research.

### 7.2 Deep Neural Recommenders

We draw a classification of successful models used in the NRS in Section 5. This information can be useful to researchers in this field, especially new researchers, to gain some knowledge and understand the guidelines on how to choose a suitable model or framework for building an NRS. For example, the Restricted Boltzmann machines (RBMs) (Salakhutdinov et al. 2007) with only two layers can be used to extract features from large news datasets using low-rank representations. The Deep Belief Network (Hu et al. 2014), a multi-layer learning architecture with a stack of RBMs, can be used to extract useful features from the news content.

There are also other DL models that can be applied in an NRS. For example, the Generative Adversarial Network (GAN) (Goodfellow et al. 2014) consisting of two competing (adversarial) neural networks (a discriminator and a generator) that run in against each other to generate new synthetic samples of data that can pass for real data. For example, GANS can be used to generate new data for an NRS with the same statistics as the training set.

There are a variety of neural networks that may be combined to produce models that are both powerful and expressive. For example, the CNNs can be used to learn feature representations from the news content and RNNs can be used for sequential user modeling. Combining AE and RNNs can capture the sequential information (through RNN) from the item content while using the lower-dimensional feature representations (through AE). These models can also be integrated with the neural attention to pick useful news recommendations.

Transfer learning can also be used to address the data sparsity problem of the NRS by transferring the knowledge from large pre-trained models to the problem of news recommendation. However, the challenge here is that the pre-trained model should be based on the news dataset. Otherwise, the noise and outliers from other unrelated datasets can be transferred into the news recommendations.

Despite the significant advances in DL theory, these methods are not without flaws. For example, DL methods demand much more data and require much more parameter adjusting than standard methods. Also, these models behave like backboxes, providing limited interpretability (due to hidden layers, weights and activation functions) and little explainability (explanation for the internal working) in the recommendation tasks.

### 7.3 Accuracy and Beyond-Accuracy Aspects, and Evaluation Protocols

We shed some light on accuracy and beyond-accuracy aspects in this survey. Accuracy is important but the quality of news recommendations cannot be improved without considering beyond-accuracy aspects. As shown in Figure 5 and Table 3 and 4, the research in beyond-accuracy aspects in NRS is limited and seems to appear trivially in recent years.

There has been some limited work in NRS that has used online evaluation and user study techniques to test. However, as seen in Tables 3 and 4, as well as Figure 6, the offline option is the most popular model evaluation protocol. Usually, the online evaluations are costly for large-scale news data could be one reason for this. One future research direction is to test these NRS models in real-world settings either by reducing the computational cost of these methods through techniques such as quantization, compression and pruning methods (Kitaev et al. 2020); or by working to manage more computational resources for in real-time experimental setup.

### 7.4 Diversity as the Key Principle in the Design of NRS

In the state-of-the-art of NRS, there is little work on the diversity aspect. Diversification in an NRS is necessary not only to keep readers engaged in the reading process, but also to keep readers from becoming trappedin filter bubbles. In order to understand why and how much diversity can be included in an NRS, academics and designers should collaborate with news organizations and social media platforms. The architecture of the news or social media website, the incorporation of the nudge theory, selective exposure, and the detection of fake news are all key aspects that should be considered while developing an NRS.

### 7.5 Diversity through Neural Attention

The neural attention can be successfully used to introduce diversity in session-based recommender systems (Nema et al. 2018). Generally, the diversity is intrinsically reflected in the users' short-term interests (Wang et al. 2018a). Under normal conditions, the attention mechanism can be used to sum the weights of the hidden layer to generate the representation vector. The issue with this approach is that if there are repeated actions in a session, the recommendations that are generated for those sessions are also similar. Therefore, it is critical to respond to users' idiosyncratic clicks during different time intervals to include diversity. A scaling weight can be assigned to the query vectors in the attention mechanism. The idea is to dampen the importance of repeated clicks and to give some weight to non-repeated users' actions using attention. So far, there is not much work in the NRS that considers including diversity in attention-based models.

### 7.6 Multi-Criteria Evaluation

There are also other aspects for evaluation that are unexplored in an NRS, such as trustworthiness (level of user trust on system), preserving privacy, efficiency (ease of the search and accessibility of the information), robustness (ability to make relevant predictions in the presence of noisy data), as well as the trade-offs among various aspects. Including these aspects in an NRS could enhance the user experience.

### 7.7 User Experience Model

There is no benchmark to evaluate user experiences in a recommender system. Also, the existing user-modeling evaluation frameworks (Konstan and Riedl 2012; Knijnenburg et al. 2012) for other recommendation domains are too expensive for an NRS. The evaluations in these frameworks are through user studies or experiments only, which is not practical in an NRS with real-time constraints. It is also a challenging task to adapt these models for the news domain. Another issue with these frameworks is that they rely only on the user study, and they do not consider any accuracy and beyond-accuracy aspects. Nonetheless, without these basic metrics, it could not offer a complete picture of the user experience. There is a need for a benchmark user modeling framework in the NRS to evaluate the experience of news readers. Such framework is not only required to provide better or enjoyable experience to the readers (as in other recommender domains), but also vital for an NRS to play its democratic, liberal and deliberative role in the community.

### 7.8 News Dataset

Our findings from Section 4 reveal that there are very few datasets in the NRS. Many of the datasets shown in Tables 2 and 3 and Figure 7 are privately owned, having been created to meet the immediate research needs of the problem to be solved. There should be more challenges, such as CLEF NEWSREEL or MIND Leaderboard, to encourage researchers to design better NRS in real-time constraints.

### 7.9 Implicit User Feedback

In an NRS, we often need implicit ratings to infer latent information from enriched user interactions. However, it could be tricky sometimes to decide whether the implicit feedback is positive or negative. For example, time spent on news articles should not always be considered as user's engagement during news reading because it could be the idle time (Agarwal and Singhal 2014). Skips from the readers are often considered as indicator of user's interest in different topics, but it could be because of repetitive news stories that force the user to skip those to find new news items (Ma et al. 2016). It is not clearly mentioned in the literature that how to find out which specific property of the system makes the recommendation uninteresting to the user. If we could devise some way to differentiate between positive and negative preferences, we can improve the quality of recommendations based on positive preferences and avoid suggesting news items to the users if these result in negative or neutral preferences.

### 7.10 Gamification

The gamification means the use of game design elements in other applications where there is no gaming context (Chou 2019). The purpose of gamification is to motivate and promote user activities. The idea of gamification has not been used in the NRS. But it can be similar to Google Guides in Google Maps. In that, the NRS can assign rewards, in the form of points, badges, avatars, leaderboards etc., to the readers based on their explicit interactions with the system. This can be a useful tool to improve user engagement and to overcome cold-start problem in the NRS.

### 7.11 Mitigating Effects of News Recommendations on Readers' Behavior

The effect of news recommendations on user behavior is one of the most overlooked area in the research of recommender systems. This topic has not attracted enough attention in the computer science field before the emergence of grave issues like fake news, deep fakes, yellow journalism (exaggerating facts or spreading rumors), ideological segregation and extremism in society due to the media war. By highlighting these problems related to the effects of news recommendations on readers' behavior in Section 6, we have presented the new research opportunities for the academic scholars to work along this direction.

So far, the solutions based on selective exposure, diversity-aware algorithms and suggestion on banning manipulative practices are not enough for two reasons: (i) they are only demonstrated on small scale experimental setup, (ii) they are based on avoidance to these techniques, which are insufficient to detect and prevent such effects from the system. The researchers in this field need to find other ways (either algorithmic or heuristic) to prevent, detect and break down those effects (filter bubbles, echo chambers) if they are prevailing. There are a few suggestions that might be useful to mitigate the effects of news recommendations. These are given below:

- ***Transparency:*** The design of news recommendation algorithms should give a much clearer view of the world as it is, not as the user wants it to be. It is no more a hidden fact that search engines such as Google use many dimensions of our online and offline behavior to determine the links that we are most likely to click from a given search[33]. In the battle to keep the news readers engaged all the time, the news recommendation algorithms are being designed in a similar way as these search engines. However, we argue that to reduce the post-algorithmic effect, we should re-design these algorithms so that they allow users to indicate their interests and then find relevant (novel, recent, important) content from diversified sources accordingly. This is much similar to introducing selective exposure and motivated information processing in the NRS.

- ***Going incognito:*** Going incognito (private mode) in browser turns off history tracking, hides cookies and logs the users out from social media sites like Google and Facebook. These social network sites transmit information about users to other websites and create echo chambers around users. In this way, the news browser is depersonalized, and a news reader receives news stories from different sites and perspectives that they would otherwise not see without incognito.

- ***Rules and regulations of recommender system's objectivity:*** User information is highly exposed during the profiling phase in the recommender systems. Although there are rules and regulations such as General Data Protection Regulation (GDPR), to protect the misuse of personal information from companies and public institutes. But when it comes to recommender systems, none of the solutions comply with these regulations. The researchers and designers of NRS need to follow these rules and regulations, not only for privacy-preserving, but to make NRS a reliable system.

---

[33] https://fortune.com/2017/06/28/gmail-google-account-ads-privacy-concerns-home-settings-policy/

### 7.12 Interdisciplinary Research

There is a need for interdisciplinary research where the expertise from both social science and computer science can be combined. The researchers may utilize the recent advancements in text analysis, representation learning and attention-based models to address the challenges specific to the news domain.

This section can only provide a partial list of some of the challenges, research directions, future opportunities and issues in the NRS. We would like to have this survey to serve as a doorway to a wealthy source of open research problems that make NRS a productive and interesting research area to work on.

## 8 Conclusion

NRS has been increasingly used in recent years to provide better suggestions to end users so that they can consume online news from various sources. There are many unique challenges associated with the NRS, most of which are inherited from the news domain. Out of these challenges, the issues related to timeliness, evolving readers' preferences over dynamically generated news, quality of news content and the effects of news recommendations on users' behavior are prominent ones. The general recommendation algorithms are insufficient to provide news recommendations since they need to be modified, varied or extended to a large extent. Recently, the DL-based solutions have addressed much of those limitations of conventional recommenders. Accuracy is considered as a standard evaluation measure to assess the quality of a recommender system. However, beyond accuracy, other aspects such as diversity, coverage, novelty, serendipity are also important to provide better user experience in an NRS. Datasets, open recommendation platforms and evaluation protocols together play a role in developing recommendation solutions in the news domain. We have covered them in this survey so that the readers can get an insight into the current research practices and may start to help develop them. Different from other survey papers, we also discuss about the effects of news recommendations on readers' behavior in this survey. Lastly, though this survey is centered around the NRS, the knowledge and insights gained from the findings of this survey can also be used to build recommender solutions for other application domains.

**Acknowledgment**: This work is partially sponsored by Natural Science and Engineering Research Council of Canada (grant 2020-04760).